\documentclass[12pt]{iopart}

\usepackage{graphicx}
\usepackage{iopams} 
\usepackage{multirow} 
\newcommand{\Part}[3]{ \frac{ \partial^{#3} #1 }{ \partial #2^{#3} }}
 
\newcommand{\Ave}[1]{\left\langle {#1} \right\rangle} 
 
\begin{document}
\title{Distribution of partition function zeros of the $\pm J$ model on the Bethe lattice}

\author{Yoshiki Matsuda$^{1,4}$, Markus M\"{u}ller$^{2}$, Hidetoshi Nishimori$^{1}$, Tomoyuki Obuchi$^{1}$ and Antonello Scardicchio$^{2,3}$}
\address{$^{1}$Department of Physics, Tokyo Institute of Technology,
Oh-okayama, Meguro-ku, Tokyo 152-8551, Japan}
\address{$^{2}$International Center for Theoretical Physics, Strada Costiera 11, 34014 Trieste, Italy}
\address{$^{3}$INFN, Sezione di Trieste, Strada Costiera 11, 34014, Trieste, Italy.}
\ead{$^{4}$matsuda@stat.phys.titech.ac.jp}

\begin{abstract}
The distribution of partition function zeros is studied for the $\pm J$ model of spin glasses on the Bethe lattice. We find a relation between the distribution of complex cavity fields and the density of zeros, which enables us to obtain the density of zeros for the infinite system size by using the cavity method. The phase boundaries thus derived from the location of the zeros are consistent with the results of direct analytical calculations. This is the first example in which the spin-glass transition is related to the distribution of zeros directly in the thermodynamical limit. We clarify how the spin-glass transition is characterized by the zeros of the partition function. It is also shown that in the spin-glass phase a continuous distribution of singularities touches the axes of real field and temperature.
\end{abstract}


\maketitle

\section{Introduction}\label{sec:intro}

The physical quantities of a spin system are completely determined by the location of the zeros of partition function on the complex parameter planes, such as the temperature and external field. This means that all the singularities of the system should be identified with zeros, and a phase transition occurs when a system parameter is driven across an accumulation point of zeros.

For ferromagnetic systems, Lee and Yang proved that the zeros on the complex $H$ plane (Lee-Yang zeros) lie on the imaginary axis. This result is the famous circle theorem \cite{Lee:52}. A ferromagnetic transition occurs when the zeros touch the real-field axis at $H = 0$, and zeros split the complex-field plane in two regions, $\Re H > 0$ and $\Re H < 0$, where $\Re$ denotes the real part. The density of zeros at the origin is directly proportional to the spontaneous magnetization in the ferromagnetic phase. As for the temperature $T$ plane (Fisher zeros), Fisher found that zeros of a pure ferromagnetic system on the square lattice in the absence of an external field  lie on the unit circle in the complex plane of $\sinh\left( 2 \beta J \right)$, where $J > 0$ is the coupling constant and $\beta = 1/T$ the inverse temperature \cite{Fisher:64}.

The Lee-Yang circle theorem is valid for random ferromagnets (where all the interactions $J_{ij}$ are greater than or equal to $0$) and all the Lee-Yang zeros lie on the imaginary-field axis. Therefore the problem is reduced to finding the density of zeros on the imaginary-field axis $g \left( \theta \right)$, which  can be calculated from the analytic continuation of the magnetization as a function of the real positive field $m \left( H \right)$, as $2 \pi g \left( \theta \right) = \Re \ m\left( 2\beta H:= i\theta \right)$ ($-\pi < \theta \le \pi$) \cite{Lee:52}. For diluted ferromagnets, several works have explored the connection between the Griffiths singularity \cite{Griffiths:69} and the density of zeros both in solvable models \cite{Bray:89, Laumann:08a} and experiments \cite{Chan:06}. These studies suggest that the density of zeros has an essential singularity on the imaginary field axis at the origin as $g\left( \theta \right) \sim \exp \left( { - A/ |\theta | } \right)$, where $A$ is a positive constant. The essential singularity at the origin, in Griffiths' interpretation \cite{Griffiths:69}, is caused by the presence of large clusters due to random fluctuations of the interactions. 

For spin glasses, the situation is more complicated. The locations of zeros are not restricted to the imaginary-field axis. Since it is difficult to treat analytically the partition functions of these systems, zeros have been studied mainly by numerical evaluation of partition functions of finite-size systems on the complex external field \cite{Ozeki:88, Matsuda:08} and temperature \cite{Bhanot:93, Saul:93, Damgaard:95} planes. The zeros were found to be distributed generally on two-dimensional areas, not only along a line as in ferromagnets. Recently a detailed study of the density of zeros on the imaginary-field axis suggested the existence of the Griffiths singularity also in spin-glass models in the paramagnetic phase \cite{Matsuda:08}.

It is therefore highly desirable to study the distribution of zeros of spin glasses in the limit of infinite system size since the complex behaviour of spin glasses should manifest itself in non-trivial distributions of zeros. We therefore investigate in this paper the relation between the distribution of zeros and the spin-glass transition in the infinite system size for the spin-glass model on the Bethe lattice. We here define the Bethe lattice as the interior part of the infinitely large Cayley tree. This definition enables us to investigate the zeros of infinite size systems without approximations (even within the spin-glass phase) using the cavity method \cite{Mezard:01}, which is perfectly suited for our purpose of studying the distribution of zeros of spin glasses. From the distribution of zeros we successfully detect the phase boundary of the spin-glass phase, defined as the line where the spin-glass susceptibility diverges, in accordance with previous studies. We also show that the system is singular everywhere as a function of the real-valued temperature or field in the spin-glass phase.

The outline of this paper is as follows. We first introduce the models and the relationship between the density of zeros and the distribution of cavity field in the next section. Then we explain the actual numerical procedures in section 3. The results are described in section 4, where distributions of zeros on the complex field and temperature planes are investigated and the phase diagrams are determined. The existence of the Griffiths singularity is also discussed. Finally we conclude this paper in section 5.

\section{Formulation}\label{sec:intro2}
In this section, the main ideas of this paper are presented. It is shown that the zeros of the partition function for the $\pm J$ model on the Bethe lattice are efficiently evaluated using the cavity method.
\subsection{Cayley tree and the cavity method}
\begin{figure}[htb]
\begin{center}
\unitlength 0.1in
\begin{picture}( 27.6000, 15.0000)(  0.4000,-15.4000)
%
\special{pn 13}%
\special{pa 116 1466}%
\special{pa 1322 828}%
\special{fp}%
\special{sh 1}%
\special{pa 1322 828}%
\special{pa 1254 842}%
\special{pa 1274 854}%
\special{pa 1272 878}%
\special{pa 1322 828}%
\special{fp}%
%
\special{pn 13}%
\special{pa 116 116}%
\special{pa 1322 746}%
\special{fp}%
\special{sh 1}%
\special{pa 1322 746}%
\special{pa 1272 696}%
\special{pa 1274 720}%
\special{pa 1254 732}%
\special{pa 1322 746}%
\special{fp}%
%
\special{pn 13}%
\special{ar 116 116 76 76  0.0000000 6.2831853}%
%
\special{pn 13}%
\special{sh 0}%
\special{ar 116 1466 76 76  0.0000000 6.2831853}%
%
\special{pn 13}%
\special{sh 0}%
\special{ar 116 116 76 76  0.0000000 6.2831853}%
%
\special{pn 13}%
\special{sh 0}%
\special{ar 1398 790 76 76  0.0000000 6.2831853}%
\put(13.6000,-10.8000){\makebox(0,0)[lb]{$i$}}%
\put(0.8000,-4.2000){\makebox(0,0)[lb]{$j$}}%
\put(7.9000,-4.3000){\makebox(0,0)[lb]{$u_j$}}%
%
\special{pn 13}%
\special{pa 1470 800}%
\special{pa 2800 800}%
\special{fp}%
\special{sh 1}%
\special{pa 2800 800}%
\special{pa 2734 780}%
\special{pa 2748 800}%
\special{pa 2734 820}%
\special{pa 2800 800}%
\special{fp}%
\put(21.4000,-10.2000){\makebox(0,0)[lb]{$u_i$}}%
\end{picture}%
 \caption{The local structure of the tree system with the coordination
 number $c=3$.}
 \label{fig:lattice}
\end{center}
\end{figure}
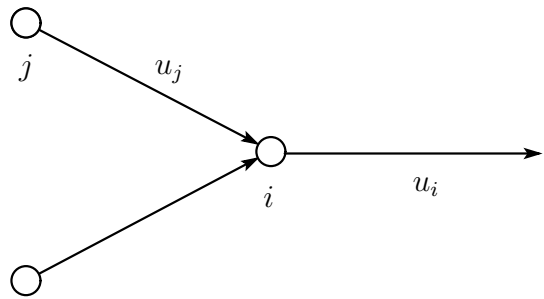

Let us consider the $\pm J$ model of spin glasses on a Cayley tree in a uniform magnetic field $H$. The Cayley tree is a cycle-free graph where each site is connected to $c$ neighbours (figure \ref{fig:lattice}). The Hamiltonian is 
\begin{equation}
\mathcal{H} =  - \sum\limits_{\left\langle {i,j} \right\rangle } {J_{ij} S_i S_j }  - H\sum\limits_i {S_i },
\end{equation}
where $\langle {i,j} \rangle$ is a nearest neighbour pair, and the value of interaction $J_{ij}$ distributes as
\begin{equation}
P\left( {J_{ij} } \right) = p\delta \left( {J_{ij}  - J} \right) + \left( {1 - p} \right)\delta \left( {J_{ij}  + J} \right)
\end{equation}
with $0 \le p \le 1$ and $J=1$.

On the Cayley tree we can efficiently calculate the partition function 
by iteratively summing over spin variables layer by layer 
from the outer boundary. The partial partition function $Z_i$, 
which represents the partition function in the absence of deeper layers than 
the site $i$, is updated as
\begin{equation}
Z_{i} = 2\cosh \left( {\beta h_i } \right)\prod\limits_{j = 1}^{c - 1} {\left[ {{{\cosh \left( {\beta J_{ij} } \right)} \over {\cosh \left( {\beta u_j(J_{ij},h_{j}) } \right)}}Z_j } \right]},
\label{pfbl}
\end{equation}
where $j$ labels $c-1$ spins on the layer previous to the current site $i$ 
(figure \ref{fig:lattice}).
The independent effects of $c-1$ spins on the previous layer have been passed to the present layer in terms of the cavity field $h_i$ and cavity biases $\{u_j\}$, 
the definitions of which are given by
\begin{eqnarray}
h_i  &=& H + \frac{1}{\beta }\sum_{j=1}^{c - 1} {\tanh ^{ - 1} \left( {\tanh \left( {\beta J_{ij} } \right)\tanh \left( {\beta h_j } \right)} \right)} \\
&=& H + \sum_{j=1}^{c-1} {u_j(J_{ij},h_{j}) }. 
\label{recursion1}
\end{eqnarray}
We hereafter assume that the function $\tanh^{-1}$ takes 
the principal branch, which restricts the value of the imaginary part of 
the bias $u_{j}$ to a range $[-\pi/\beta,\pi/\beta]$.
From now on we assume the imaginary part of all the fields defined modulo $\pi/\beta$ and the principal branch of the $\tanh^{-1}$ is considered. The reader should keep this in mind, as we will not burden the notation with this condition explicitly.

At the final step of this procedure, we consider the contribution from 
the central site of the Cayley tree, and the partition function of the whole system 
$Z^{(c)}$ is obtained as 
\begin{equation}\label{pfcent}
Z^{(c)}  = 2\cosh \left( {\beta h^{(c)} } \right)\prod\limits_{j = 1}^c {\left[ {{{\cosh \left( {\beta J_{ij} } \right)} \over {\cosh \left( {\beta u_j(J_{ij},h_{j}) } \right)}}Z_j } \right]},\label{pfc} 
\end{equation}
and
\begin{equation}\label{cfcent}
h^{(c)} = H + \sum_{j=1}^{c} {u_j(J_{ij},h_{j}) },\label{hc}
\end{equation}
where the superscript $\left.^{(c)}\right.$ represents the central site. 

To consider the typical behaviour of the Cayley tree (where we assume uncorrelated, typical boundary conditions, to be specified explicitly below), we perform the average over the quenched randomness, 
which introduces the distribution of the cavity field at the $l$th layer $P_{H,\beta}^{l}$. The update rule of $P_{H,\beta}^{l}$ is given by
\begin{equation}
P_{H,\beta}^{l+1} \left( {h} \right) =  {\int { \left[ \delta (h - H - \sum\limits_{j = 1}^{c - 1} {u_j\left( h_j \right) } ) \right]_J} \prod\limits_{j = 1}^{c - 1} {P_{H,\beta}^{l} \left( {h_j } \right)dh_j }  }  ,
\end{equation}
where $\left[ \cdots \right]_J$ denotes the average over the random interactions $J_{ij}$. 
In the thermodynamic limit, and if the distributions $P_{H,\beta}^{l}$ converge as $l\to\infty$ to $P_{H,\beta} = \lim_{l \rightarrow \infty} P_{H,\beta}^{l}$, the limiting distribution satisfies the following self-consistent equation
\begin{equation}
P_{H,\beta} \left( {h} \right) =  {\int { \left[ \delta (h - H - \sum\limits_{j = 1}^{c - 1} {u_j\left( h_j \right) } ) \right]_J}\prod\limits_{j = 1}^{c - 1} {P_{H,\beta} \left( {h_j } \right)dh_j }  }\label{eq:convergentCFD} .
\end{equation}
The existence of a unique solution of this equation is 
shown rigorously in \cite{Chayes:86} for real
values of the cavity field. Although this proof is not directly applicable to our case of
complex cavity field, we observed numerically that equation~(\ref{eq:convergentCFD}) has a solution for $p < 1$. The special case of pure ferromagnet ($p=1$)  in a purely imaginary external field is discussed in detail in Appendix A.

\subsection{Definition of the Bethe lattice}
The Cayley tree is peculiar because the number of surface sites is comparable to the total number of sites and hence the contribution from the surface cannot be neglected. 
This property is inconvenient for studying the bulk of the system. 
Thus the Bethe lattice is often considered instead. There are two different definitions of the Bethe lattice, and we have to clearly distinguish them \cite{Obuchi:09}.
\begin{enumerate}
 \item The interior part of the Cayley tree: A lattice consisting of the 
interior part of an infinite-size Cayley tree with uncorrelated boundary conditions. 
Alternatively, we can define this lattice as a finite  
Cayley tree, the boundary conditions of which are given by the convergent 
cavity field distribution $P_{H,\beta}=\lim_{l \rightarrow \infty} P_{H,\beta}^l(h)$ 
of the infinite-size Cayley tree.

 \item The regular random graph (RRG): A randomly generated graph under the constraint of a fixed
connectivity $c$. Since there exist many cycles, we cannot exactly treat the finite system size. 
In the limit $N \to \infty$ under appropriate conditions (and outside the spin-glass phase), however, the contribution coming from the loops is expected to become negligible
and the problem can be solved exactly by the cavity method.
\end{enumerate}

Both definitions have advantages and disadvantages, and typically the phenomena that occur in one set-up have a correspondence in the other. The model (i) and its phase diagram have been discussed in detail in \cite{Chayes:86}, while for (ii) and its relations with the replica theory we refer the reader to reference~\cite{Mezard:01}.

In this paper, we refer to model (i) (with typical boundary conditions which are not correlated among each other or with the couplings in the interior) 
and call it the Bethe lattice.
 Although definition (ii) is useful in that it has no ambiguity about the boundary conditions, it is incompatible with the formulation of the partition function zeros that uses equation~(\ref{pfbl}). 
On the other hand, according to definition (i), we can easily define a finite-size system of the Bethe lattice 
and look at its zeros in the formalism of the cavity method.


\subsection{The partition function zeros of the Bethe lattice}
Let us first consider the zeros of the partition function with respect to 
the external field $H$, the Lee-Yang 
zeros \cite{Lee:52}. 
It is known that these zeros are related to the magnetization of the system.
In general, the partition function of an Ising system in an external field is expressed as a two-variable polynomial,
\begin{equation}\label{pf}
Z\left(H, \beta \right) = e^{N\beta H} e^{N_b\beta J} \sum\limits_{M = 0}^N {\sum\limits_{E = 0}^{N_b} {\Omega \left( {E,M} \right)e^{-2E\beta J} e^{-2M\beta H} } } ,
\end{equation}
where $N$ is the number of sites and $N_b$ is the number of interactions. We can write the above equation with fixed temperature as
\begin{equation}
Z\left( H \right) = \xi e^{ N\beta H} \prod\limits_{i}^N {\left( {e^{-2\beta H}  - e^{-2\beta H_i} } \right)} ,\label{pfpolynomial}
\end{equation}
where $\xi $ is a constant (to be ignored hereafter). Equation (\ref{pfpolynomial}) means that the partition function is a polynomial of degree $N$ in the fugacity $e^{-2 \beta H}$, and there are $N$ roots on the complex $H$ plane. Taking the logarithm and dividing by $N$, we find
\begin{equation}
 - \beta f\left( H \right) = \beta H + \int\!\!\!\int {d^2 H'g \left( {H'} \right)\log \left( {e^{ - 2\beta H}  - e^{ - 2\beta H' } } \right)}.
\end{equation}
where $g_H$ is defined as the density of zeros on the complex $H$ plane. The complex magnetization $m\left( H \right) = -\partial f/\partial H$ is expressed as
\begin{equation}\label{mag(H)}
m\left( H \right) = 1 + \int\!\!\!\int {d^2 H'g \left( {H'} \right){2 \over {e^{2\beta \left( {H - H'} \right)}  - 1}}} .
\end{equation}
We can express the density of zeros in terms of the magnetization by an infinitesimal closed line integral,
\begin{equation}
\label{g2(H)}
g(H)= \frac{\beta}{2\pi i} \lim_{r\to 0} \frac{1}{\pi r^2}\oint_{|H-H'|=r} m(H') dH'
\end{equation}
as is easily seen from the representation (\ref{mag(H)}). Indeed the line integral picks up all poles within the circle, and all poles have the residue $1/\beta$.

For the Bethe lattice, due to the uniformity of the system, 
the disorder averaged complex magnetization is given by
\begin{equation}\label{mcav}
m = \int\!\!\!\int {d^2 h^{(c)} P_{H,\beta}^{(c)}\left( {h^{(c)}} \right)\tanh \left( \beta h^{(c)} \right) },
\end{equation}
where the distribution of the central complex field $P_{H,\beta}^{(c)}$ 
is calculated from the convergent distribution $P_{H,\beta}$ as
\begin{equation}\label{fdcent}
P_{H,\beta}^{(c)} \left( {h^{(c)}} \right) =  {\int\!\!\!\int {\left[\delta (h^{(c)} - H - \sum\limits_{j = 1}^{c} {u_j\left( h_j \right) } ) \right]_J\prod\limits_{j = 1}^{c} {P_{H,\beta} \left( {h_j } \right)d^2h_j } } }  .
\end{equation}
Inserting the equation~(\ref{mcav}) into equation~(\ref{g2(H)}), 
we obtain
\begin{eqnarray}
\label{gh}
\label{eq:PH1}
g(H) &=& \frac{\beta}{2\pi i}
\lim_{r\to 0} 
\frac{1}{\pi r^2}\oint_{|H-H'|=r} dH' \int\!\!\!\int {d^2 h^{(c)} P_{H'}^{(c)}\left( {h^{(c) }} \right)\tanh \left( \beta h^{(c)}\left( H' \right) \right) } \nonumber\\
&=& 
\lim_{r\to 0} 
\frac{1}{\pi r^2}\int\!\!\!\int_{|H-H'| \le r} d^2 H' \int\!\!\!\int d^2 h^{(c)} P_{H'}^{(c)}\left( h^{(c) } \right)\delta \left( \beta h^{\left( c \right)} \left( H' \right) - \frac{\pi i}{2} \right)  \nonumber\\
&=& 
\lim_{r\to 0} 
\frac{1}{\pi r^2}\int\!\!\!\int_{|H-H^*|\leq r}\!\!\!\!\!\!\!\!\!\! d^2H' {P_{H'}^{(c)}\left(\pi i /2\beta  \right)}  \nonumber\\
&=& {P_H^{(c)}\left(\pi i /2\beta  \right)}\label{eq:PH1}.
\end{eqnarray}
In the second line, we used the residue theorem under the condition that the radius $r$ is sufficiently small. 
This result connects the density of zeros at $H$ with the density of cavity fields at fixed external field $H$. If we iterate the cavity field population numerically for a given pair of complex values $\left( H, \beta \right)$, the distribution function of the cavity field yields the density of zeros in the $H$-plane through equation~(\ref{gh}).

The relation (\ref{gh}) can also be interpreted as follows. 
The whole partition function of the Bethe lattice $Z^{(c)}$ 
is formally the same as that of the Cayley tree (\ref{pfcent}), the difference only being whether or not we use the limiting distribution $P_{H,\beta}(h)$.
This implies that the equation of zeros $Z^{(c)}=0$ 
becomes identical to\footnote{Note that equation~(\ref{pfcent}) may 
seem to diverge when 
the factor $\cosh{\beta u_j }$ becomes $0$, but this is not the case. 
The reason is that the condition 
$\cosh{\beta u_j }=0$ is always accompanied by $Z_{j}=0$ and  
$(\cosh{\beta u_j })^{-1}Z_{j}$ yields a finite value.} 
\begin{equation}
2\cosh \left( {\beta h^{(c)} }\right)=0 
\Rightarrow
\beta h^{(c)}=\frac{\pi}{2} i.\label{cosh=0}
\end{equation}

Equations (\ref{gh}) and (\ref{cosh=0}) show that the complex support of the zeros of the partition function can be obtained from the knowledge of the distribution of the field acting on the central spin.  This exact relation represents an important advantage of considering the Bethe lattice as the interior part of the infinite-size Cayley tree. It is one of the main results of the present paper.

We now turn our attention to the density of zeros as a function of the complex temperature.
The condition (\ref{cosh=0}) for the vanishing of the partition function is still valid for complex values of temperatures. In order to find the correct density one should follow steps analogous to those which lead from (\ref{pfpolynomial}) to (\ref{eq:PH1}), by treating $\partial f/\partial \beta$ instead of $\partial f/\partial H$. 
We have however not followed this route here. 
Instead we contented ourselves with detecting the support of the zeros, i.e., the location on the temperature plane where $g \left( \beta\right)$ is non-zero. This can be performed by  
using the fact that 
the value of the density on the field plane 
at a complex temperature $\beta = \beta '$, 
$\left. g\left( H = H' \right) \right|_{\beta = \beta'}$,
should be proportional to the density on the temperature plane 
in a field $H=H'$,
$\left. g\left( \beta = \beta' \right)\right|_{H = H'}$.
This means that the equality 
$\left. g\left( H = H' \right) \right|_{\beta = \beta'} = C\left( \beta ' \right) \left. g\left( \beta = \beta' \right)\right|_{H = H'}$ holds, 
where $C(\beta')$ is a normalization factor (to be dropped hereafter)
\footnote{We can accept this equality by considering fact that  
the zeros are located in the four-dimensional complex temperature-field space, 
and the two-dimensional distributions $g \left( \beta \right)$ with real $H$ and $g \left( H \right)$ with real $\beta$ are just two-dimensional cross sections of the same distribution function in four dimensions.}. 
This equality is sufficient to understand the phase diagram in terms of the locations of zeros, while a quantitative knowledge of the density of zeros on the temperature plane cannot be obtained due to the nontrivial normalization factor $C(\beta')$.
Anyway we do not discuss quantitatively the value of density on the temperature plane in the present work.

\section{Numerical procedures}
Hereafter, we discuss the density of zeros $g ( \hat H )$ on the $\hat H = 2 \beta H$ plane since $H$ always appears in this combination. A hat on a variable, like $\hat H$, will mean the same quantity multiplied by $2 \beta$. The location of the zeros on the complex temperature plane will also be analyzed.

For the spin-glass model, the zeros on the complex field plane are not restricted to the imaginary field axis but a finite fraction spreads into the complex plane. The remaining fraction still lies on the imaginary field axis. The total density of zeros on the field plane thus naturally splits into two parts,
\begin{equation}
g(\hat H) = \delta(\Re \hat H)g_1(\theta) + g_2 (\hat H),
\end{equation}
where $\theta = \Im \hat H$ ($-\pi \le \theta \le \pi$) is the imaginary part of $\hat H$. The one-dimensional measure $g_1$ is restricted to the imaginary axis and the two-dimensional measure $g_2$ is continuously distributed on the complex plane.

We base our analysis mostly on equation~(\ref{gh}). However, it is not sufficient for our purpose because it only yields the two-dimensional part $g_2$. In fact, the cavity field distribution always spreads in the two-dimensional $h$-plane due to the initial conditions that we choose for the Bethe spin glass, as explained later. It is not possible to obtain the one-dimensional part $g_1$ together with the two-dimensional $g_2$ from equation~(\ref{gh})~\footnote{For non-frustrated systems, we can obtain $g_1$ from the relation (\ref{gh}) (see \ref{densitypuref}).}. However we can calculate $g_1$ by a different method using the fact that the one-dimensional density on the imaginary field axis corresponds to the jump of the real part of the complex magnetization as $H$ crosses the imaginary axis. In other words, $2 \pi g_1 \left( \theta \right) = \Re \ m( \hat H= i\theta +0^+ )$ \cite{Lee:52}. The analogy with electrostatics developed in reference~\cite{Lee:52} allows us to apply this formula even when $g_2 \ne 0$. Therefore we can calculate $g_1$ from equation~(\ref{mcav}).

We now obtain the distribution of $P_{\hat H}^{(c)}\:( {\hat h^{(c)} =2 \beta h^{(c)}} )$ for complex field $H$ and temperature $T$. Let us explain the algorithms to obtain $P_{\hat H}^{(c)}\:( {\hat h^{(c)} } )$ and estimate the density of zeros. We write the recursion relation (\ref{recursion1}) as
\begin{equation}\label{recursion2}
u_j  = {1 \over \beta }\tanh ^{ - 1} \left( {\tanh \left( {\beta J_{ij} } \right)\tanh \left( {\beta \left( {H + \sum\limits_{k = 1}^{c - 1} {u_k } } \right)} \right)} \right),
\end{equation}
where the interaction $J_{ij}$ connects the current site $j$ with the next site $i$, and $k$ labels the previous sites. The central cavity field is
\begin{equation}\label{recursion3}
\hat h^{(c)}  = \hat H + 2 \beta \sum\limits_{j = 1}^c {u_j } .
\end{equation}
Our numerical solution for the probability densities $\pi_H \left( u \right)$ and $P_{\hat H}^{(c)}( \hat h^{(c)} )$ is based on the method of population dynamics. We represent the distribution functions $\pi_H \left( u \right)$ and $P_{\hat H}^{(c)} \: ( \hat h^{(c)} )$ in terms of a large number of variables $\left\{ u_j \right\}$ and $ \: \{ \hat h^{(c)} \}$, whose distributions are supposed to follow the respective probability distributions. The elements of these sets are updated by randomly choosing $u_i$ and $J_{ij}$ and following equations~(\ref{recursion2}) and (\ref{recursion3}). The initial condition of the cavity bias, which correspond to the boundary condition of the Cayley tree, is taken to be $u_j = \lim_{H \rightarrow \infty} u_j = J_{ij} \left( \in \mathbb{R} \right)$. These initial fields are uncorrelated random variables, which introduce frustration into the system. Note that in general the value of $u_j$ has both real and imaginary parts because the value of the next generation is calculated from equations~(\ref{recursion2}) and (\ref{recursion3}) with complex $\beta$ and $H$. The distribution of $\hat h^{(c)}$ is calculated from the Monte Carlo (dynamical) average of the population in the two dimensional complex plane, after the dynamics has converged to a limiting 
distribution.

The actual steps of the algorithm are as follows;
\begin{enumerate}
 \item Set the initial probability distribution of $u_j$ as $u_j = J_{ij}$.
 \item Update the sets of $\left\{ u_j \right\}$ and $ \: \{ \hat h^{(c)} \}$ with the recursion relations (\ref{recursion2}) and (\ref{recursion3}) by randomly choosing $u_k$ and $u_j$ out of the set $\left\{ u_j \right\}$ until $P_H^{(c)} \:( \hat h^{(c)} )$ converges.
  \item Estimate the density of zeros by
\begin{eqnarray}
g_1 \left( \theta, T \right) &=& {1 \over {2\pi }} \Re \ m( \hat H = i \theta+0^+ ) \nonumber\\
&=& {1 \over {2\pi }} \Re \lim_{  \hat H_R  \rightarrow 0^+}  \left. \int\!\!\!\int {d^2 \hat h^{(c)} P_{\hat H}^{(c)}\left( { \hat h^{(c)}} \right)\tanh \left(\hat h^{(c)} /2 \right) } \right|_{\hat H = i \theta + \hat H_R}  \label{densityg1}
\\
g_2 \: ( \hat H, T ) &=&  P_{\hat H}^{(c)} ( \hat h^{(c)} = \pi i ) 
 \label{density}
 \end{eqnarray}
  for given $\hat H$ and $T$. The one-dimensional density $g_1$ represents a density of zeros under a pure-imaginary field while the two-dimensional density $g_2$ is defined on the whole complex planes. 
\end{enumerate}
For simplicity only the Bethe lattice with connectivity $c=3$ has been studied. We have chosen $N_{\rm pop} = 10^6$ representative points in the population dynamics and have performed at least $5000 N_{\rm pop}$ cavity iterations until the population converges. The data were collected from the average of additional $5000 N_{\rm pop}$ iterations  after the initial $5000 N_{\rm pop}$ (or more) iterations.  These conditions have been used throughout this work.

\section{Distribution of zeros for the Bethe spin glass}
\subsection{Zeros on the complex field plane}\label{oncomplezfp}
\begin{figure}[tb]
\begin{minipage}{0.5\hsize}
\includegraphics[width=1.00\linewidth]{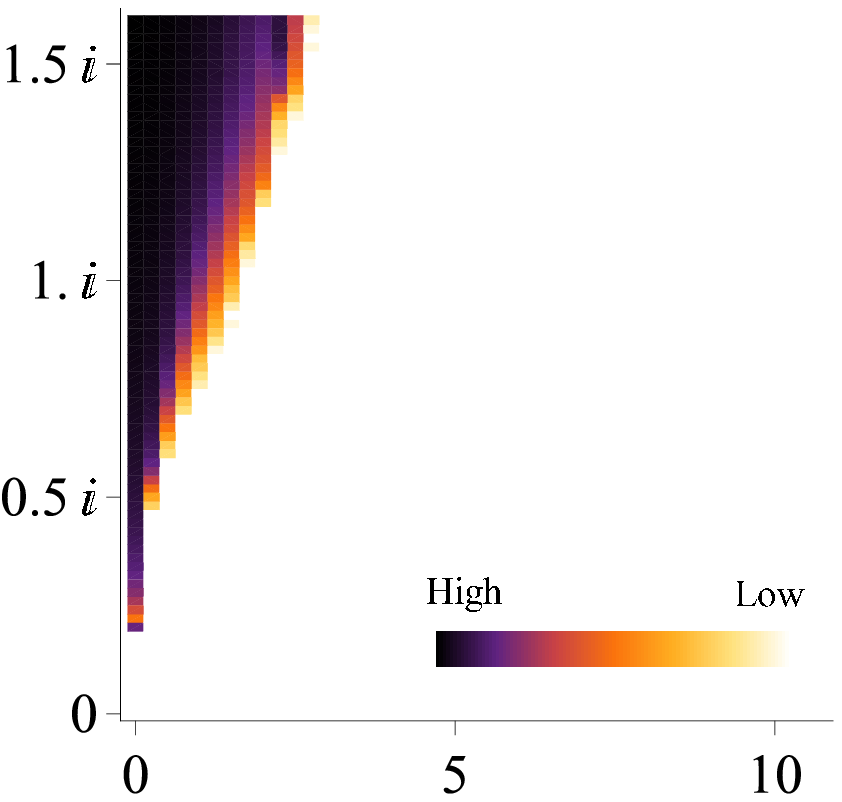}
\end{minipage}
\begin{minipage}{0.5\hsize}
\includegraphics[width=1.00\linewidth]{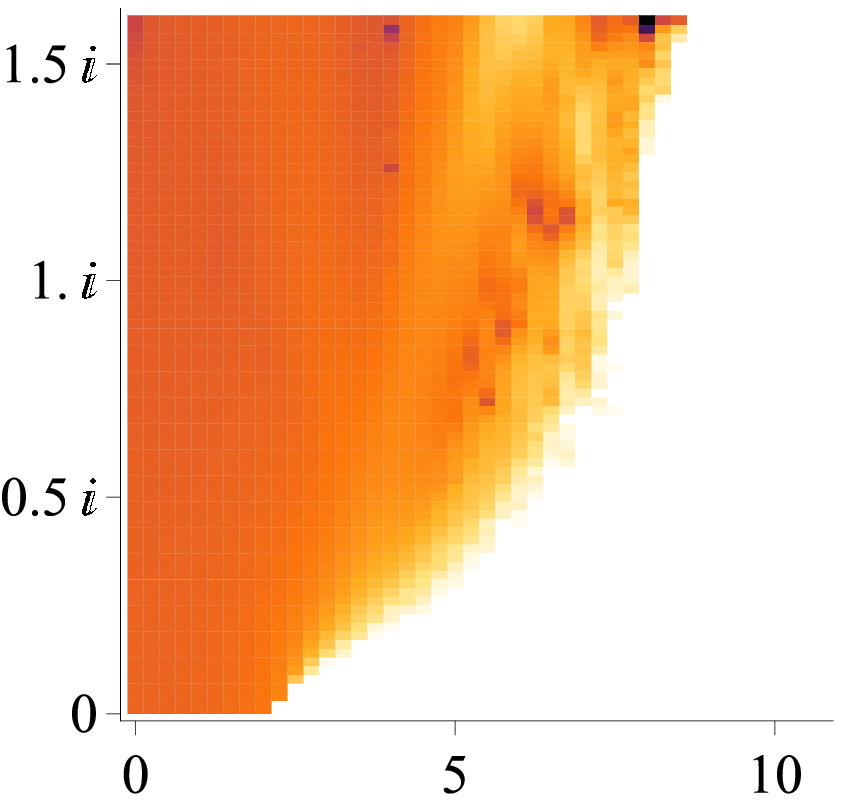}
\end{minipage}
\caption{
Distribution of zeros on the complex $\hat H$ plane with $p=0.5$ at $T = 1.43$ (paramagnet, left panel) and $T=0.5$ (spin-glass, right panel). Densities are coloured in a logarithmic scale. On the right panel, zeros with finite real part touch the real axis, whereas $g_1$ on the imaginary axis vanishes to the numerical precision.
\label{fig:p05parasg}}
\end{figure}

First, we show the density of zeros on the complex $\hat H$ plane for real temperature. Figure \ref{fig:p05parasg} is the distributions of zeros on the complex $\hat H $ plane with probability $p = 0.5$ at temperature $T = 1.43 > T_{SG} = 1/\tanh ^{ - 1} \left( {1/\sqrt {c - 1} } \right) $ (left) and $T=0.5 < T_{SG}$ (right). The complex plane has been split into cells by dividing the real axis from $\Re \: ( \hat H ) = 0$ to $12$ with an increment $0.25$ and the imaginary axis from $\Im \:( \hat H ) = 0.02$ to $\pi/2$ with an increment of $0.02$.
The density outside this range is omitted, since this is sufficient to see zeros near the real axis which are essential for critical phenomena. Both $g_1$ and $g_2$ are plotted in the same figure and coloured in a logarithmic scale; a black dot shows a very high density.

The left panel of figure \ref{fig:p05parasg} ($T=1.43$) is for the paramagnetic phase. Neither $g_1$ nor $g_2$ has a finite value on the real axis and there is thus no phase transition as a function of the real field. The right panel of figure \ref{fig:p05parasg} is in the spin-glass phase ($T=0.5$), where zeros reach the real axis at and away from the origin. Thus, there is a phase transition at some $ H= H_c (>0) \in \mathbb{R} $ on the Bethe lattice, where $H_c$ is the point where the density $g_2$ changes its value from zero to non-zero along the real axis. Zeros touch the real axis all the interval between $H=0$ and $H=H_c$. This suggests that the free energy is non-analytic as a function of $H$ below $H_c$. The one-dimensional density $g_1$ vanishes to the numerical precision for both temperatures in figure \ref{fig:p05parasg}. The two-dimensional part $g_2$ has finite values on the imaginary axis. It is likely that the edge of $g_2$ (the point where $g_2 
 \ne 0$ on the imaginary axis and is closest to the origin) touches the origin as the temperature is decreased to $T=T_{SG}$. At lower $T$ the density of the zeros approaches the real axis also away from the imaginary axis, up to a point $H_c(T)$ on the real axis. This defines a line of spin-glass transitions in the $H$-$T$ plane, as discussed below. 

\begin{figure}[tb]
\begin{minipage}{0.5\hsize}
\includegraphics[width=1.00\linewidth]{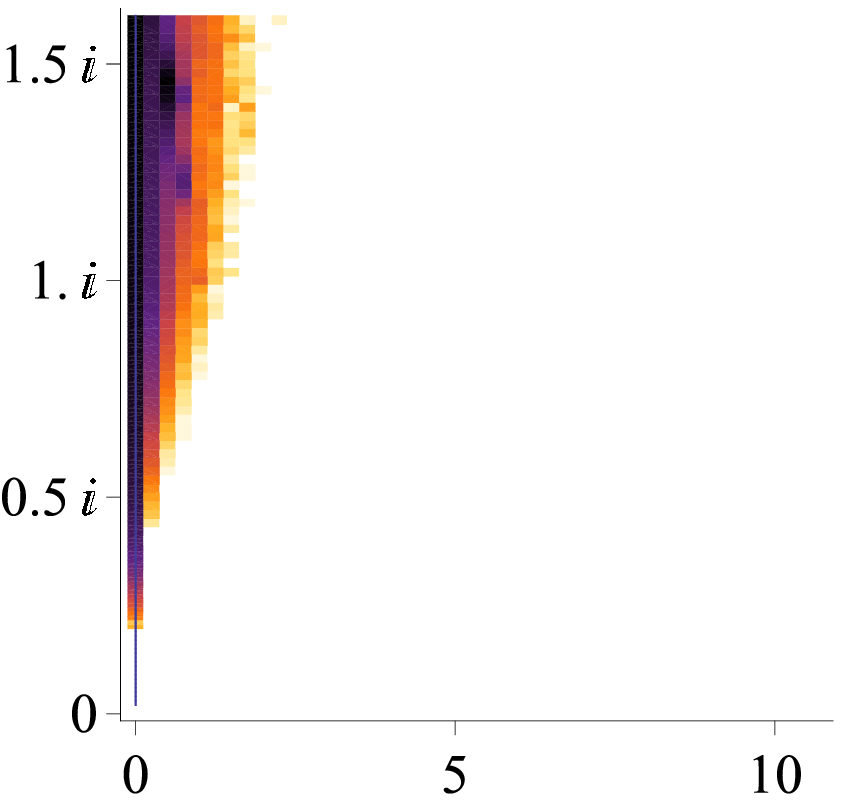}
\end{minipage}
\begin{minipage}{0.5\hsize}
\includegraphics[width=1.00\linewidth]{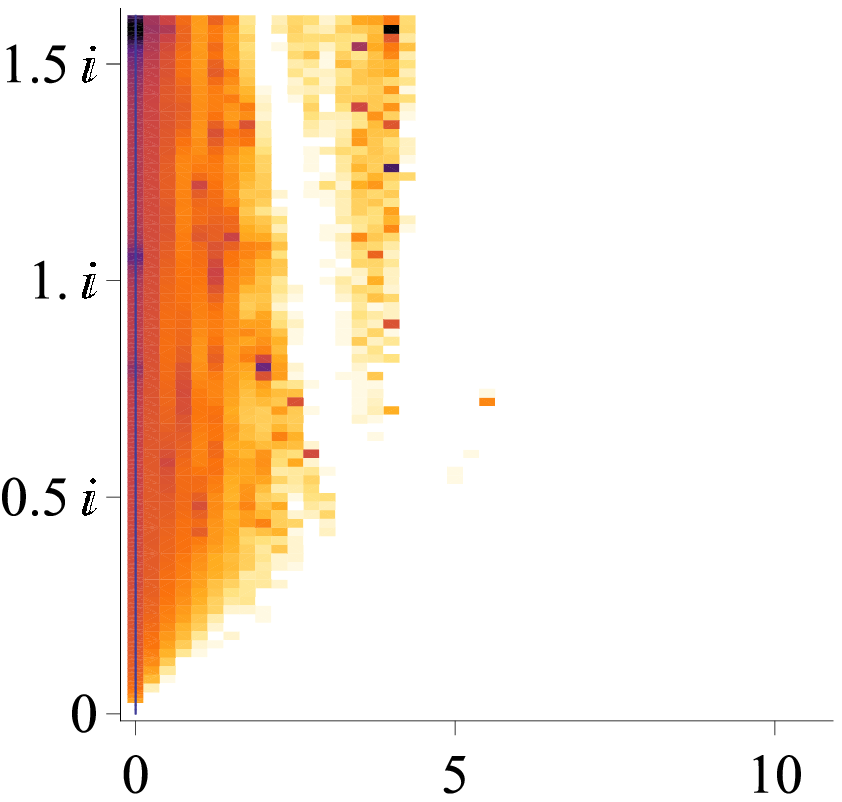}
\end{minipage}
\caption{
Distribution of zeros on the complex $\hat H$ plane with $p=0.9$ at $T = 1.5$ (left: para) and $T=0.5$ (right: ferro). Only the one-dimensional zeros on the imaginary axis, $g_1$ (thin lines), touch the origin at low temperature. 
\label{fig:p09paraferro}}
\end{figure}

In the ferromagnetic phase (the right part of figure \ref{fig:p09paraferro}), on the other hand, only $g_1$ touches the origin but $g_2$ does not. In marked contrast to the left panel of figure \ref{fig:p05parasg}, the one-dimensional density has a finite value on the imaginary axis away from the origin on the left panel of figure \ref{fig:p09paraferro}. This is an important difference of ferromagnetic and spin-glass transitions seen from the distribution of Lee-Yang zeros. More details on the transition points will be discussed in later sections.

\subsection{Zeros on the complex temperature plane}
Using complex values of the temperature in equation~(\ref{recursion2}), we can also obtain information on the support of Fisher zeros, using equation~(\ref{density}). This information is sufficient for our purpose of detecting phase transitions from the point of view of zeros.

\begin{figure}[tb]
\begin{minipage}{0.5\hsize}
\includegraphics[width=1.00\linewidth]{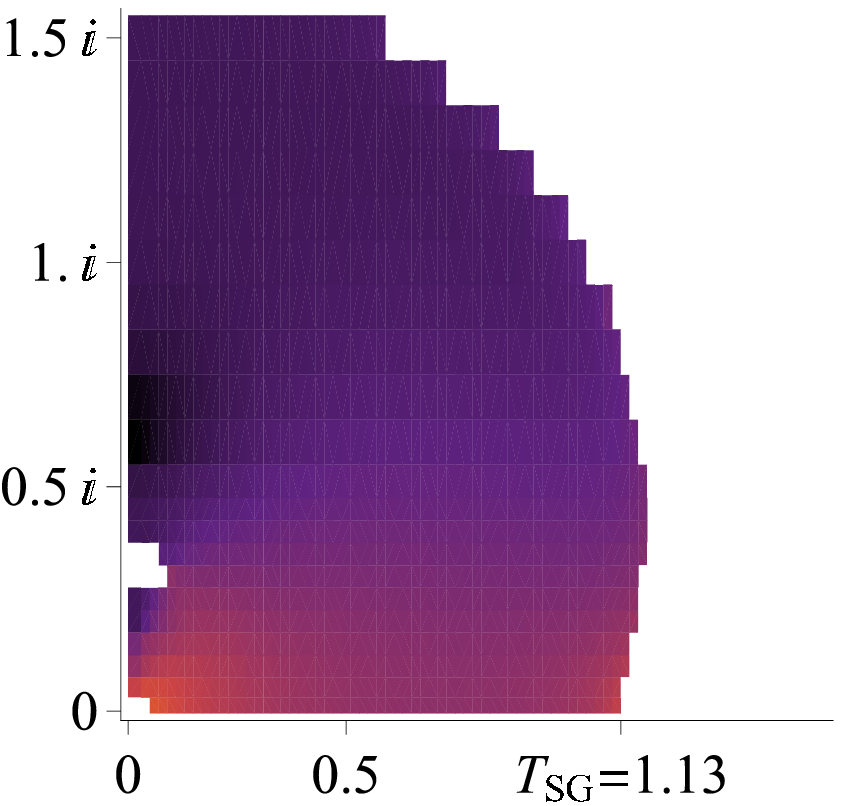}
\end{minipage}
\begin{minipage}{0.5\hsize}
\includegraphics[width=1.00\linewidth]{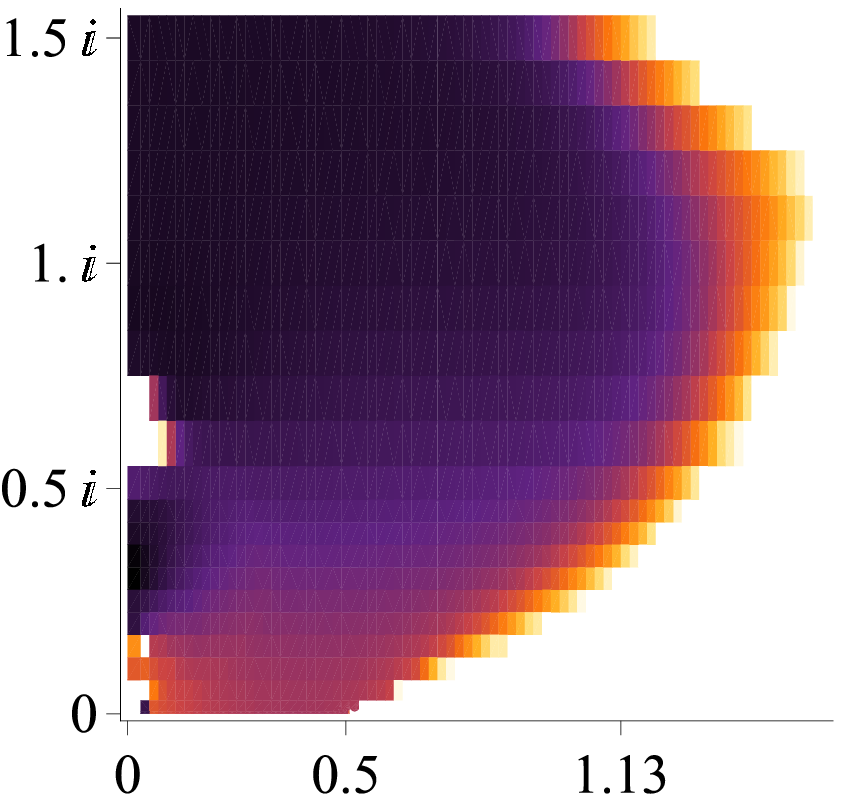}
\end{minipage}
\caption{
Distribution of zeros on the complex $T$ plane with $p=0.5$ and $H=0$ (left) and $H=0.5$ (right). Zeros touch the real axis below the critical temperature $T_{SG} \simeq 1.13$ (left) and $0.5$ (right). The apparent absence of zeros near the origin may be due to numerical rounding errors of $\tanh \beta$.
\label{fig:p05parasg_tp}}
\end{figure}

Figure \ref{fig:p05parasg_tp} shows the distribution of zeros on the temperature plane at $p=0.5$ with $H = 0$ (left) and $H = 0.5$ (right). The (two-dimensional) zeros also touch the real temperature axis below the spin-glass transition temperature. This confirms that the spin-glass transition differs from the ordinary phase transition where zeros touch the real axis only at the transition temperature. This may be interpreted as the system staying critical at all temperatures below the transition temperature. It may also be taken as a signature of \emph{temperature chaos}, i.e. the instability of the randomly frozen spin configurations in the spin-glass phase with respect to arbitrarily small temperature changes. In accordance with the right panel of figure \ref{fig:p05parasg}, the right panel of figure \ref{fig:p05parasg_tp} shows that the spin-glass phase persists under a weak field ($H=0.5$). The critical temperatures appearing in figure \ref{fig:p05parasg_tp} are consistent with the phase diagram shown in the next subsection.

\begin{figure}[tb]
\begin{minipage}{0.5\hsize}
\includegraphics[width=1.00\linewidth]{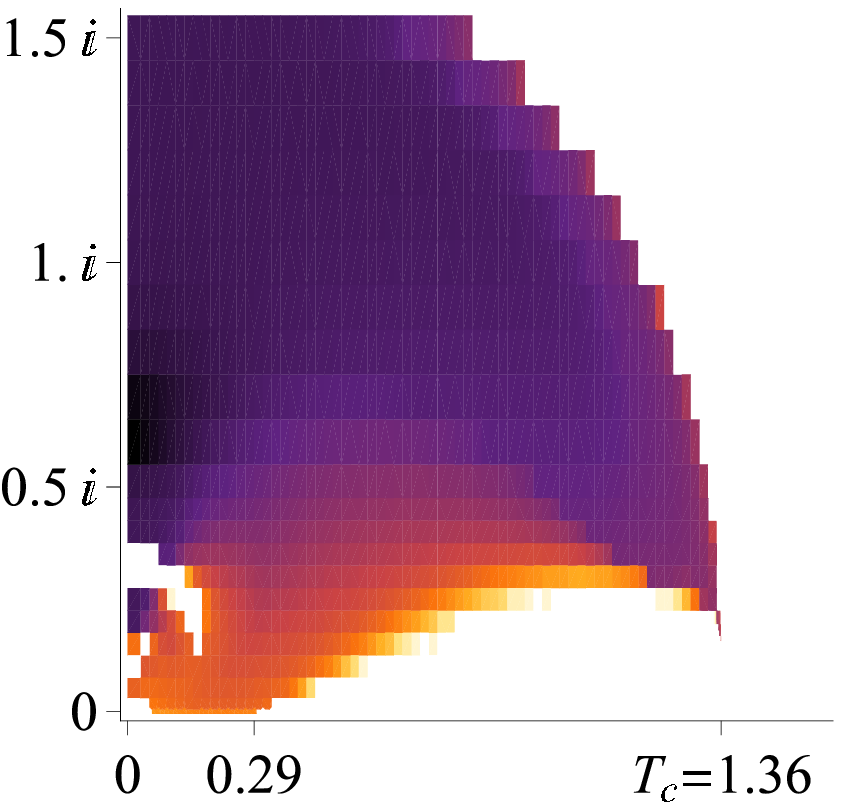}
\end{minipage}
\begin{minipage}{0.5\hsize}
\includegraphics[width=1.00\linewidth]{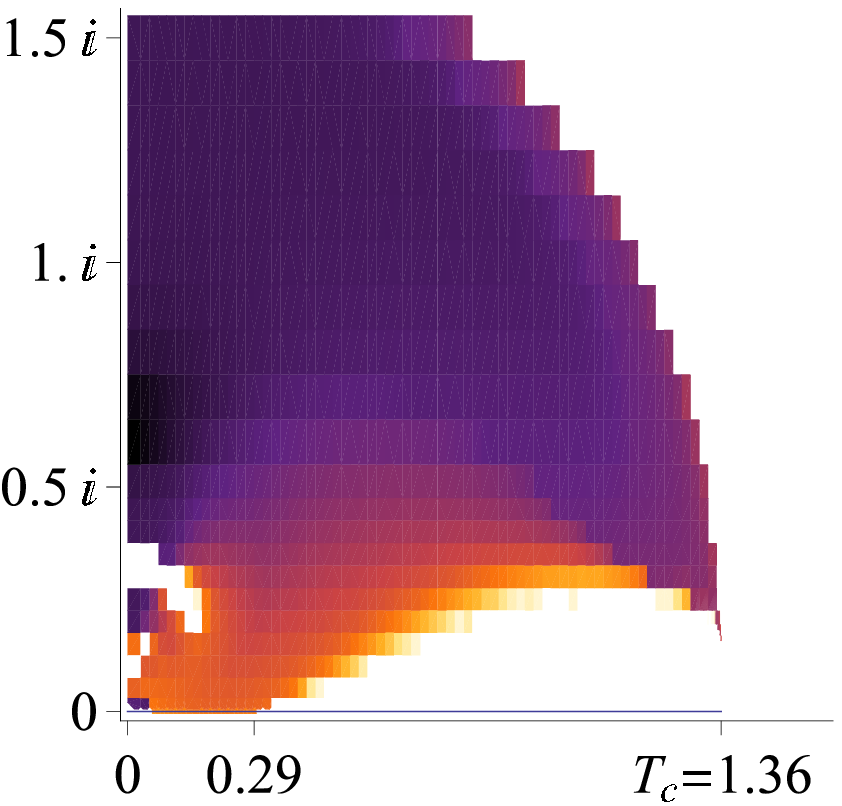}
\end{minipage}
\caption{
Distribution of zeros on the complex $T$ plane for $p=0.9$ and $\hat H=0$ (left) and $\hat H=10^{-4}i$ (right). The zeros approach the real axis at $T=T_c$, where $T_c \simeq 1.36$ is the ferromagnetic critical temperature. In addition, zeros reach the real axis also in the low temperature region. The second critical temperature corresponds to the spin-glass transition $T_{SG}$ shown in figure \ref{fig:pdpt}. On the right panel, a weak pure-imaginary field immediately makes apparent the presence of ferromagnetic order below a critical temperature $T_c$: a one-dimensional distribution of zeros lies on the real axis in the right panel.
\label{fig:p09parasg_tp}}
\end{figure}

The distributions at $p$ close to 1 are also interesting. Figure \ref{fig:p09parasg_tp} shows the density at $p=0.9$ with $\hat H=0$ (left) and $\hat H=10^{-4}i$ (right). The transition temperature between  paramagnetic and ferromagnetic phases is known to be $T_c \simeq 1.36$. It seems that the zeros distribute on an almost one-dimensional curve near $T_c$ which most likely touches the real axis at $T = T_c$. Since our method of equation~(\ref{density}) calculates the two-dimensional density of zeros, a one-dimensional density is hard to identify with a high precision. On the right panel of figure \ref{fig:p09parasg_tp}, a weak field $\hat H=10^{-4}i$ is added, to test for a spontaneous magnetization: zeros have finite densities along the real axis below $T_c$, since $g_1$ is finite under a pure imaginary field in the ferromagnetic phase. 

On the other hand, the two-dimensional distribution $g_2$ touches the real axis again in the low temperature region. Below $T_{SG} \simeq 0.29$, the zeros approach the real axis on the whole interval $0<T<T_{SG}$, similarly as in the spin-glass phase shown in figure \ref{fig:p05parasg_tp}. Indeed, this second critical temperature corresponds to the spin-glass transition, as shown below.

\subsection{Phase diagram}
\begin{figure}[tb]
\begin{minipage}{0.525\hsize}
\includegraphics[width=1.15\linewidth]{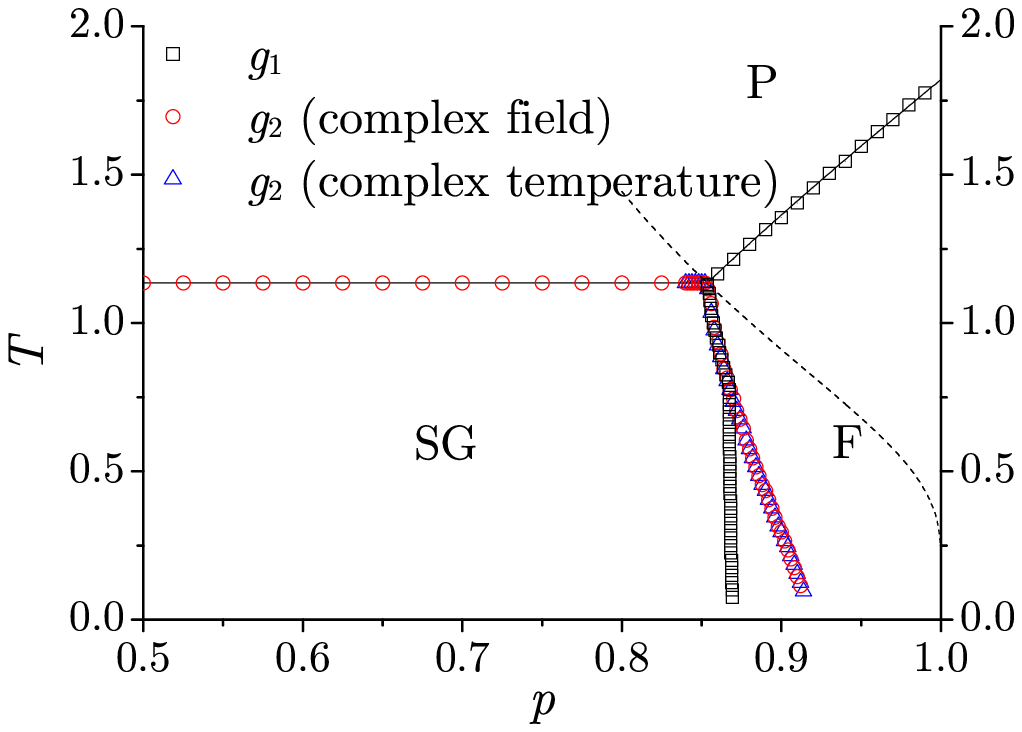}
\end{minipage}
\begin{minipage}{0.5\hsize}
\includegraphics[width=1.1\linewidth]{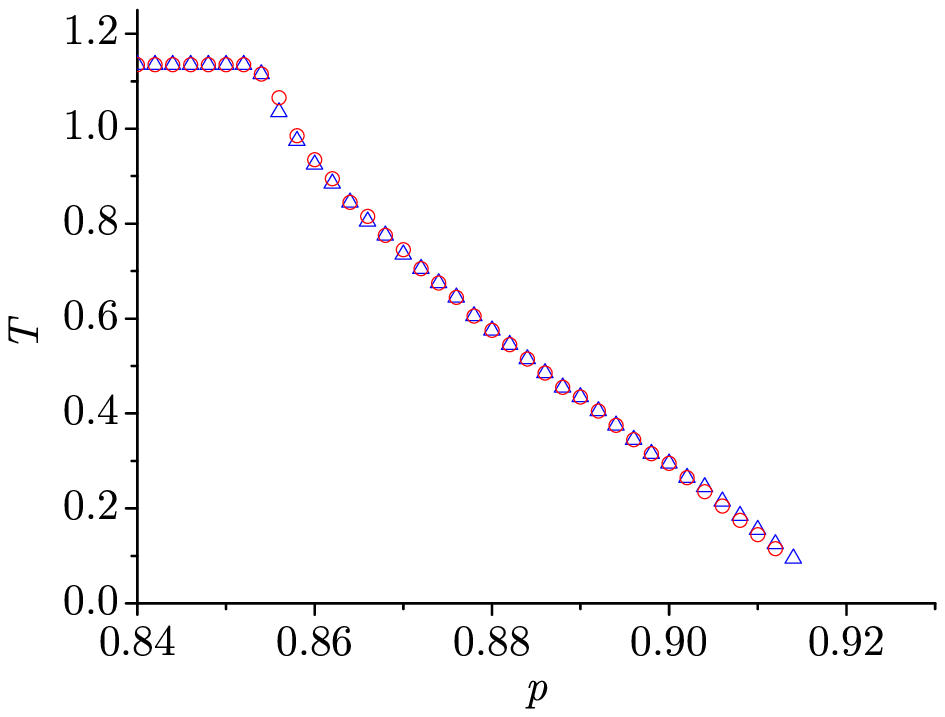}
\end{minipage}
\caption{
Phase diagrams on the $p$-$T$ plane from the density of zeros. The solid lines are analytical results as given as $T_c = 1/\tanh^{-1}\left[{ {1/ \left( 4p-2\right) } }\right]$ (para-ferro boundary) and $T_{SG} = 1/\tanh^{-1}\left[{ {1/ \sqrt{2} } }\right]$ (para-spin glass boundary) \cite{Thouless:86,Carlson:90,Carlson:90a,Kabashima:03}. Left: Circles and triangles are determined by the two-dimensional density of zeros $g_2$ on the complex field plane and temperature plane, respectively, which corresponds to spin-glass critical temperatures. Squares are phase boundaries calculated by the one-dimensional density of zeros $g_1$, which indicates the onset of the ferromagnetic order. Immediately below triangles there is a phase which is both magnetized and glassy. The dotted line denotes the Nishimori line \cite{Nishimori:81}, on which the multicritical point (where three phases merge) is located. Right: Blow up of the field-induced critical temperatures on the left panel (marked in circles and triangles) between $p=0.84$ and $p=0.93$.

\label{fig:pdpt}}
\end{figure}

Based on the above observations, we investigate the phase diagram on the $p$-$T$ and $T$-$H$ planes with real $T$ and $H$. Graphical representations shown so far suggest that there are two types of transitions where one- and two-dimensional distributions of zeros reach the real axis separately. In order to determine the two transition points, we added a very small imaginary field $\hat H = 10^{-4} i$. Since zeros on the imaginary axis in the complex $\hat H$ plane reach the real axis before those away from the imaginary axis as the temperature is decreased, we can restrict to the imaginary $H$ axis and ask for the highest temperature for which a non-zero density, $g_1$ and $g_2$ exists at the origin of the imaginary axis ($\hat H = 10^{-4} i$). In contrast, it is difficult to determine the transition points caused by one-dimensional zeros on the complex $T$ plane in the absence of a field as shown in the left panel of figure \ref{fig:p09parasg_tp}. The transition temperature $T_c \simeq 1.36
 $ derived from the first method of the complex $\hat H = 10^{-4} i$ is close to the point in the left panel of figure \ref{fig:p09parasg_tp} where the almost one-dimensional distribution is likely to touch the real axis. Therefore we mainly used the method of complex field to identify the phase boundary (the left panel of figure \ref{fig:pdpt}).

Our result agrees well with the exact phase boundary \cite{Thouless:86,Carlson:90,Carlson:90a,Kabashima:03}, $T_c = 1/\tanh^{-1}\left[{ {1/ \left( 4p-2\right) } }\right]$ (between para and ferro phases) for $p>p_c=(2+\sqrt{2})/4$ and $T_{SG} = 1/\tanh^{-1}\left[{ {1/ \sqrt{2} } }\right]$ (between para and spin-glass phases) for $p<p_c$. It is expected that the one-dimensional distribution $g_1$ determines the ferromagnetic temperature $T_c$ and the two-dimensional distribution $g_2$ determines the spin-glass transition temperature $T_{SG}$. We discuss this hypothesis in detail below. 

\subsubsection{Behaviour of the two-dimensional distribution of zeros\\}\label{sec:twodim}

Below the line drawn in circles in figure \ref{fig:pdpt}, the ordered phase is expected to be stable under an external field as shown in figure \ref{fig:p05parasg}. This line is also determined from the zeros on the temperature plane. We set the temperature as $T = T_R+ 10^{-3}i$ with $H=0$, where $T_R$ is real, and we assume the highest $T_R$ having a non-zero density as the critical temperature (triangles in figure \ref{fig:pdpt}). The two results agree well with each other.

The boundary between the ferromagnetic and spin-glass phases is harder to determine. As one sees in figure \ref{fig:pdpt} (the left panel), the temperature where the one-dimensional distribution $g_1$ ceases to touch the real axis is lower than the temperature where the two-dimensional distribution $g_2$ starts touching the real axis (for biases $p\simeq 0.85$ to $0.92$). This indicates that there is a phase where the system is still spontaneously magnetized, but also glassy (see also \cite{Chayes:86}). It is shown in Appendix B that the spin-glass susceptibility $\chi_{SG}$ diverges along a line which coincides with the touching points of $g_2$, marked in circles and triangles in figure \ref{fig:pdpt} . This implies that our method using the approach of the two-dimensional distribution $g_2$ to the real axis correctly detects the onset of the spin-glass phase. Indeed the proximity of zeros to the real axis at $H=0$ implies the divergence of some (higher order) susceptibility. 

\begin{figure}[tb]
\begin{minipage}{0.5\hsize}
\includegraphics[width=1\linewidth]{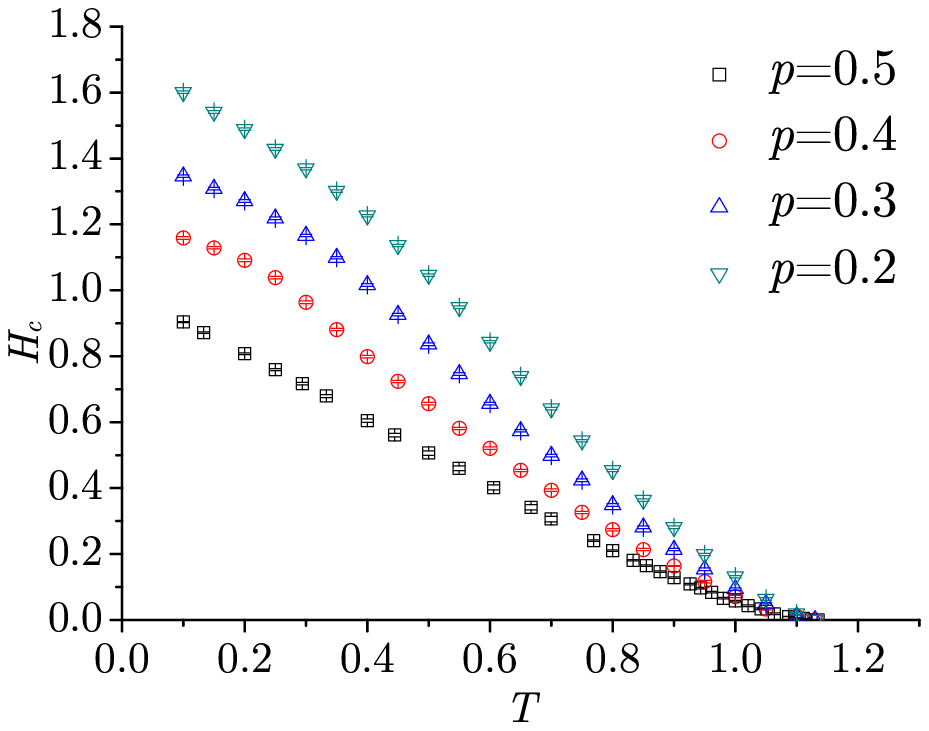}
\end{minipage}
\begin{minipage}{0.5\hsize}
\includegraphics[width=1\linewidth]{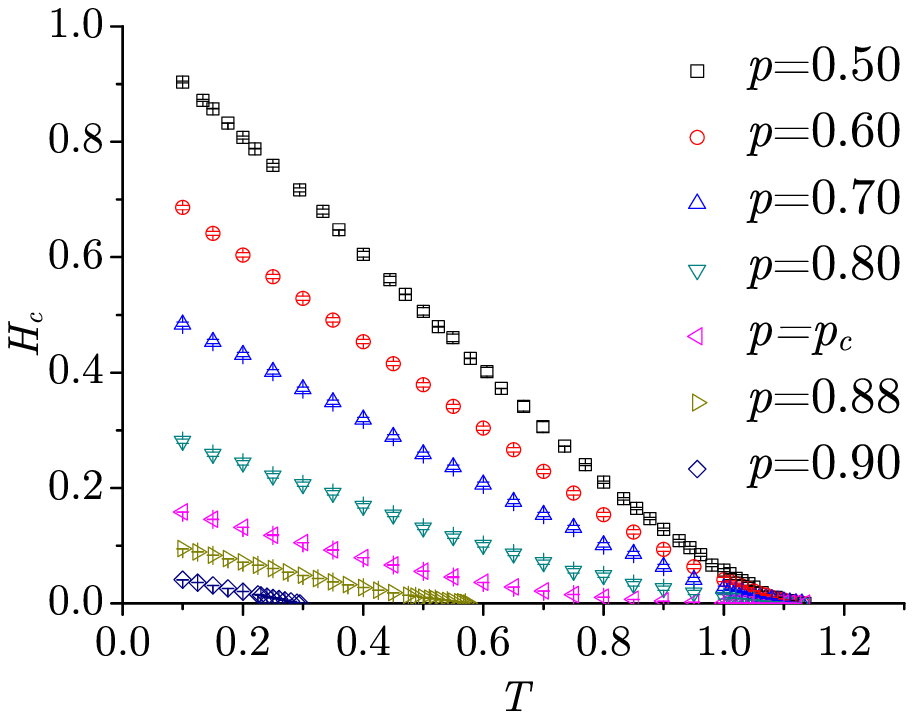}
\end{minipage}
\begin{minipage}{0.5\hsize}
\includegraphics[width=1\linewidth]{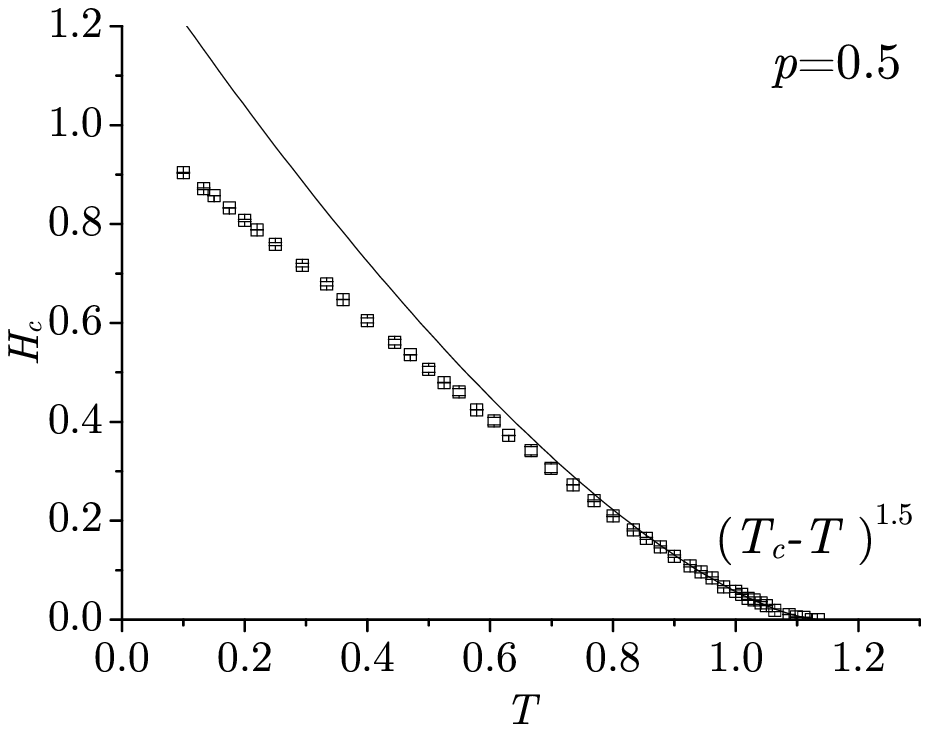}
\end{minipage}
\begin{minipage}{0.5\hsize}
\caption{
Spin-glass transition lines in the $T$-$H$ plane (corresponding to the AT line in the SK model). The bottom panel is for $p = 0.50$. The boundary at $p = 0.5$ rises from $T=T_c$ as $H_c  \propto \left( {T_c  - T} \right)^{1.5}$. The exponent $1.5$ agrees with that of the AT line for the SK model. $p_c = \left( 2+\sqrt{2} \right)/4 \simeq 0.854$ is for the multicritical point.
\label{fig:pdHT}}
\end{minipage}
\end{figure}

We next show the phase diagram on the $T$-$H$ plane with fixed $p$ (figure \ref{fig:pdHT}). We added a complex external filed as $\hat H = \hat H_R + 10^{-4}i$ and we assumed that the critical value of $H_c$ is the maximum value of $\hat H_R / 2\beta = H$ having a non-zero (two-dimensional) density $g_2$. The top panels of figure \ref{fig:pdHT} are for the temperature dependence of the critical line for various $p$. 
For all biases $p$, $H_c$ smoothly rises from $H=0$ at $T=T_c \left( H=0 \right)$ for $p \le p_c = \left( 2+\sqrt{2} \right)/4 \simeq 0.854$, which is for the multicritical point on the Nishimori line \cite{Nishimori:81} where the three phases merge. At low temperature, it seems that $H_c$ approaches a finite value at $T=0$. The bottom panel is for $p = 0.5$, and the points well fit to $H_c  \propto \left( {T_c  - T} \right)^{1.5}$. The exponent $1.5$ is the same as in the AT line \cite{Almeida:78} of the Sherrington-Kirkpatrick model \cite{Sherrington:75}. The existence of the AT line on the Bethe lattice (regular random graph) was predicted in reference~\cite{Thouless:86} where the AT line should behave like $\left( T_c -T \right)^{3/2}$ in the vicinity of $T=T_c$, and was checked by the population method in references~\cite{Pagnani:03, Jorg:08}.

The agreement of the exponent $1.5$ with the behaviour of the AT line might be interpreted as an indication that the phase below the phase boundary is the spin-glass phase with replica symmetry breaking (RSB) \cite{M'ezard:87, Fisher:91, Nishimori:01}. Indeed, it is suggested that the spin-glass phase in the regular random graph has the full RSB~\cite{Castellani:05, Krzakala:05}. However, we should be careful because we have not directly seen the instability of a replica-symmetric solution in the Bethe lattice of our definition. The analysis of reference~\cite{Chayes:86} shows that in the spin-glass phase a system of two replicas on the Cayley tree exhibits a diverging susceptibility with respect to an infinitesimal repulsion between the replicas. However, at the same time the cavity field distribution remains essentially identical to the replica symmetric approximation for the regular random graph; model (ii). 
Further studies of this phase from different viewpoints may be necessary to fully clarify its nature.

\subsection{Griffiths singularity}
The density of zeros on the imaginary field axis is closely related to the Griffiths singularity in the diluted ferromagnet \cite{Griffiths:69}. The same is expected in spin glasses \cite{Matsuda:08}. The Griffiths singularity is expected to manifest itself, if it is present, in the form of an essential singularity of the density of zeros upon approaching the origin along the imaginary field axis \cite{Bray:89, Laumann:08a, Chan:06, Matsuda:08}. If such a tail is present (which is very difficult to detect numerically), its touching of the real axis would indicate the onset of a Griffith phase, but not the phase transition to a ferromagnetically ordered or glassy state. In the presence of a Griffith phase, our criterion for the detection of phase transitions should thus strictly speaking be refined to the condition that a "substantial density of zeros" (which grows at least like a power law with $\Im{H}$ away from the real axis) touches the real axis.

The above discussion suggests to study in more detail the form of the one-dimensional density of zeros on the imaginary axis as an indicator of possible Griffiths singularity for the Bethe spin glass.

\begin{figure}[tb]
\begin{minipage}{0.5\hsize}
\includegraphics[width=1\linewidth]{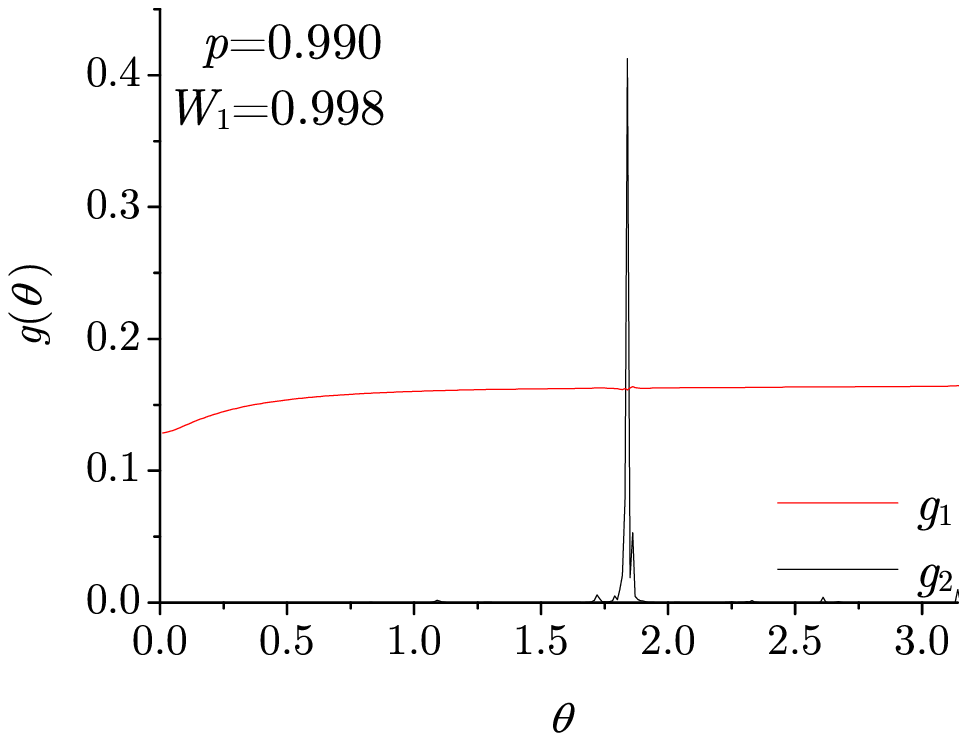}
\end{minipage}
\begin{minipage}{0.5\hsize}
\includegraphics[width=1\linewidth]{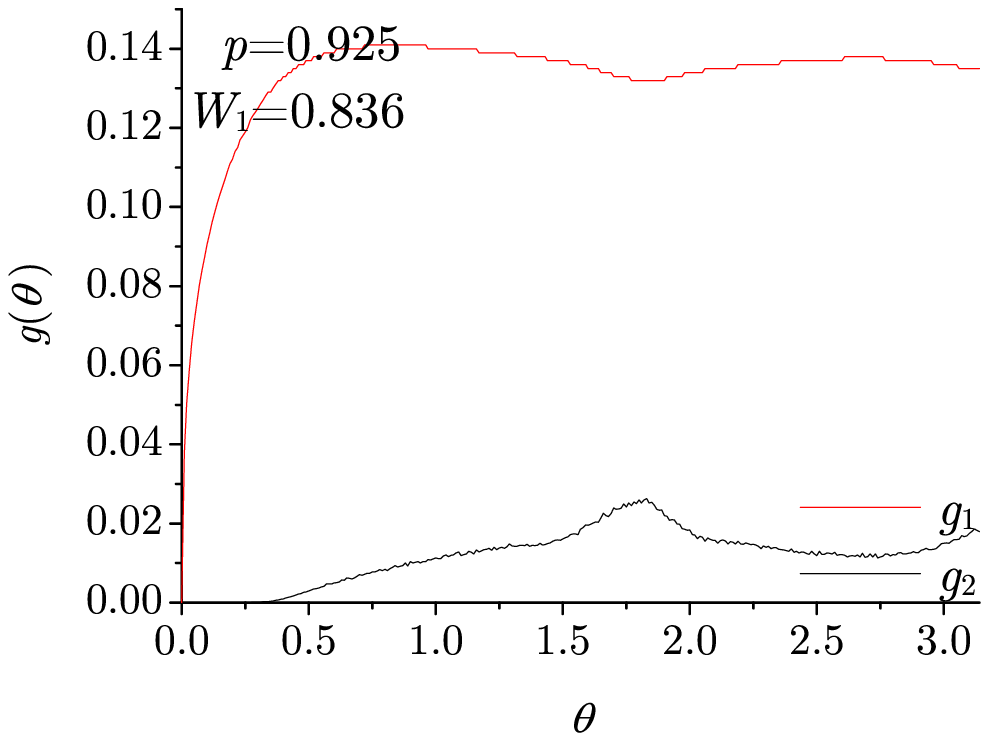}
\end{minipage}
\begin{minipage}{0.5\hsize}
\includegraphics[width=1\linewidth]{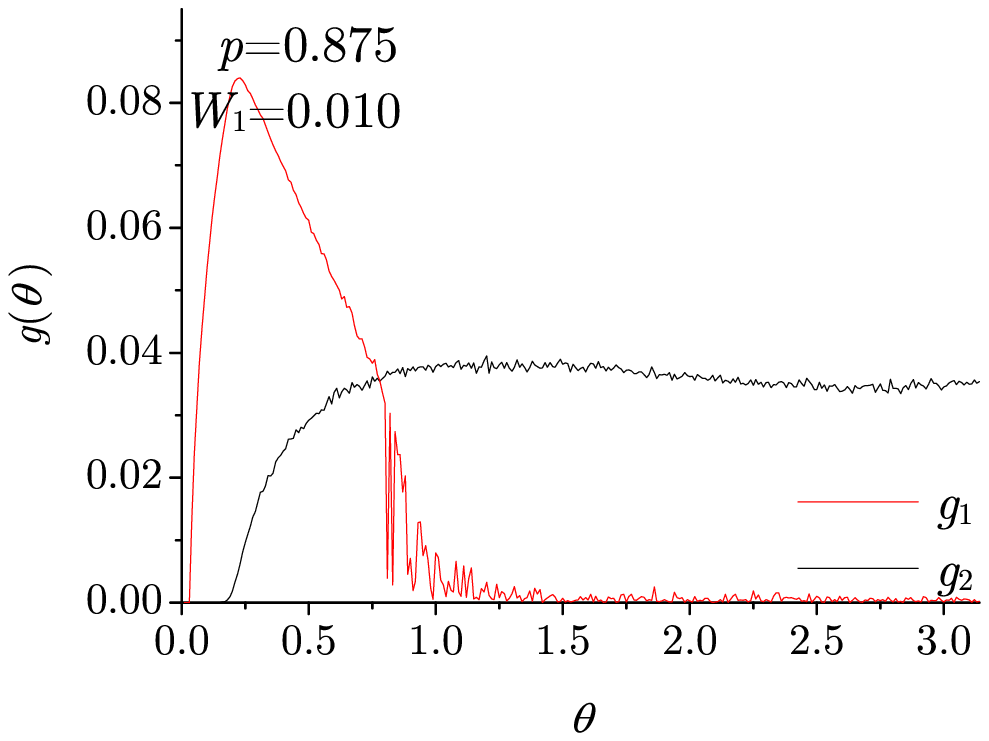}
\end{minipage}
\begin{minipage}{0.5\hsize}
\includegraphics[width=1\linewidth]{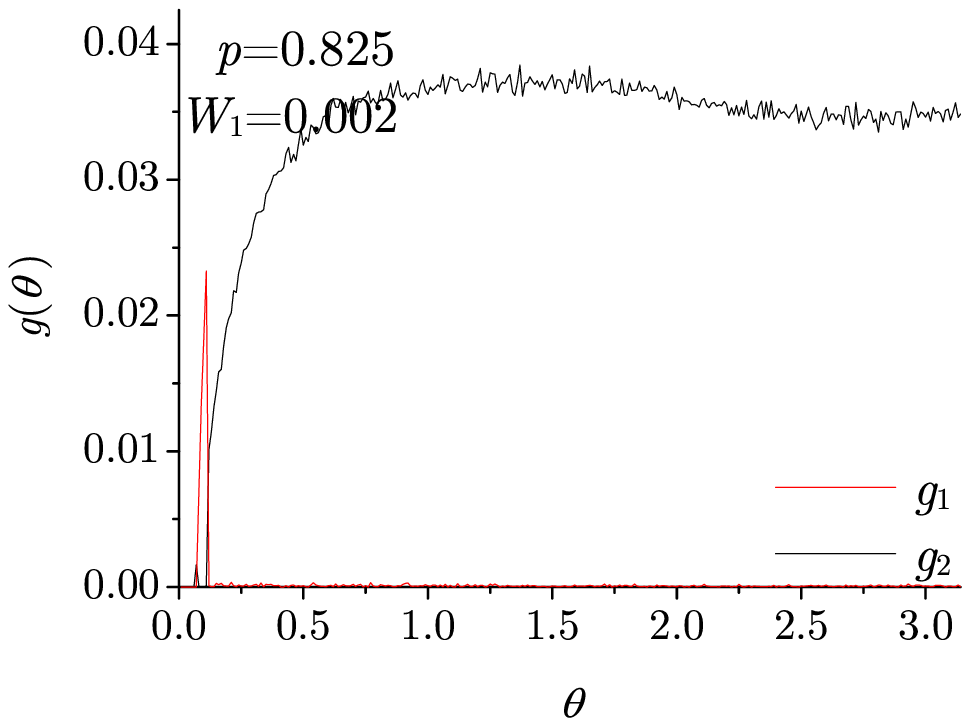}
\end{minipage}
\caption{
One- and two-dimensional density of zeros on the imaginary field axis for high $p$ at $T=1.5$. The top panels are for the ferromagnetic phase and the bottom for the paramagnetic phase. The contribution of the one-dimensional density $g_1$ decreases with decreasing $p$ whereas $g_2$ increases.
\label{fig:g1g2img}}
\end{figure}
For the Bethe spin glass our numerics does not show detectable signs of a Griffiths singularity above the spin-glass phase, our estimated $g_1$ being zero within numerical precision in the paramagnetic phase above the spin-glass phase, as shown in figure \ref{fig:p05parasg}. Above the ferromagnetic phase, it seems that $g_1$ is finite above some $\theta_0~(>0)$, while there seems to be a jump between $g_1=0 \left( \theta < \theta _0 \right)$ and $g_1 > 0 \left( \theta > \theta _0 \right)$.

Figure \ref{fig:g1g2img} shows the one- and two-dimensional densities of zeros on the imaginary axis at fixed $T=1.5$ with various $p$. The top panels correspond to biases $p$ which are in the ferromagnetic phase while the bottom panels correspond to lower biases, in the paramagnetic phase. At high $p$, the $g_1$ part is dominant because the line integrals 
\begin{equation}
W_1  = \int_{ - \pi }^\pi  {g_1 \left( \theta  \right)d\theta } 
\end{equation}
for $p=0.99$ and $0.925$ are nearly equal to $1$, where $W_1  + W_2  = 1$ with $W_2  = \int\!\!\!\int {d\hat Hg_2 \:( {\hat H} \:)d\hat H}$ should hold due to the normalization. On the other hand, $W_1$ rapidly decreases with decreasing $p$, and $g_1$ vanishes for large $\theta$. In the left-bottom panel of figure \ref{fig:g1g2img}, $g_1$ is seen to decrease when $g_2$ becomes finite around $\theta = 1.0$. This behaviour suggests that one-dimensional zeros are buried under large cloud of two-dimensional zeros far away from the real axis. Finally, for $p<p_c$ the one-dimensional $g_1$ seems to vanish altogether above the spin-glass phase as shown in figure \ref{fig:p05parasg}.

It is difficult to draw a definite conclusion about Griffiths singularities due to the limit to detect very small but non-vanishing values of $g_1$ near the origin. However, our numerics points toward the possible absence of a Griffiths phase in the spin glass on the Bethe lattice, contrary to what happens for finite-dimensional systems and diluted ferromagnets. A more detailed study will be required to settle this question. 

\section{Conclusion}
We have investigated the distribution of the partition function zeros for the $\pm J$ spin-glass model on the Bethe lattice. An important relation to connect the density of zeros and the density of cavity fields for complex field and temperature is found, which enables us to treat the zeros in the limit of infinite system size. The densities are split into one- and two-dimensional parts, where the one-dimensional density is defined on the imaginary axis of the complex field plane and the two-dimensional density spreads over the complex plane. We investigated the phase diagram of the Bethe spin glass by estimating the transition points where the one- or two-dimensional densities begin to have finite values in the immediate vicinity of the real axis. Our results agree well with the analytically exact phase diagram. The one-dimensional density determines the boundary between the paramagnetic and ferromagnetic phases since the one-dimensional density is directly related to the spontaneous magnetization. The two-dimensional density determines the boundary of the spin-glass phase on the $p$-$T$ and $H$-$T$ planes. This phase boundary corresponds to the line where the spin-glass susceptibility diverges as evaluated in Appendix B. Below the spin-glass transition point, the two-dimensional density of zeros continuously touches the real field and temperature axes, which may be related to chaotic behaviour of the system as a function of the field and temperature in the spin-glass phase. We have shown, by looking at the locations of the partition function zeros, that the system has a non-analytic free energy for all $T$ below the spin-glass temperature $T_c$.

We have not observed any evidence for the existence of a Griffiths phase in our numerics. If such a phase is indeed absent, it implies that the Bethe lattice behaves distinctly from finite-dimensional spin glasses. Further careful studies are required to understand the origin of differences between Bethe and Bravais lattices.

\ack
This work was partially supported by CREST, JST and by the Grant-in-Aid for Scientific Research on the Priority Area `Deepening and Expansion of Statistical Mechanical Informatics' by the Ministry of Education, Culture, Sports, Science and Technology of Japan. YM and TO are grateful for the financial support provided through the Japan Society for the Promotion of Science (JSPS) Research Fellowship for Young Scientists Program.
\appendix

\def\thesection{Appendix \Alph{section}}

\section{Zeros of the pure ferromagnetic system}\label{densitypuref}
\begin{figure}[tbp]
\begin{minipage}{0.5\hsize}
\includegraphics[width=1\linewidth]{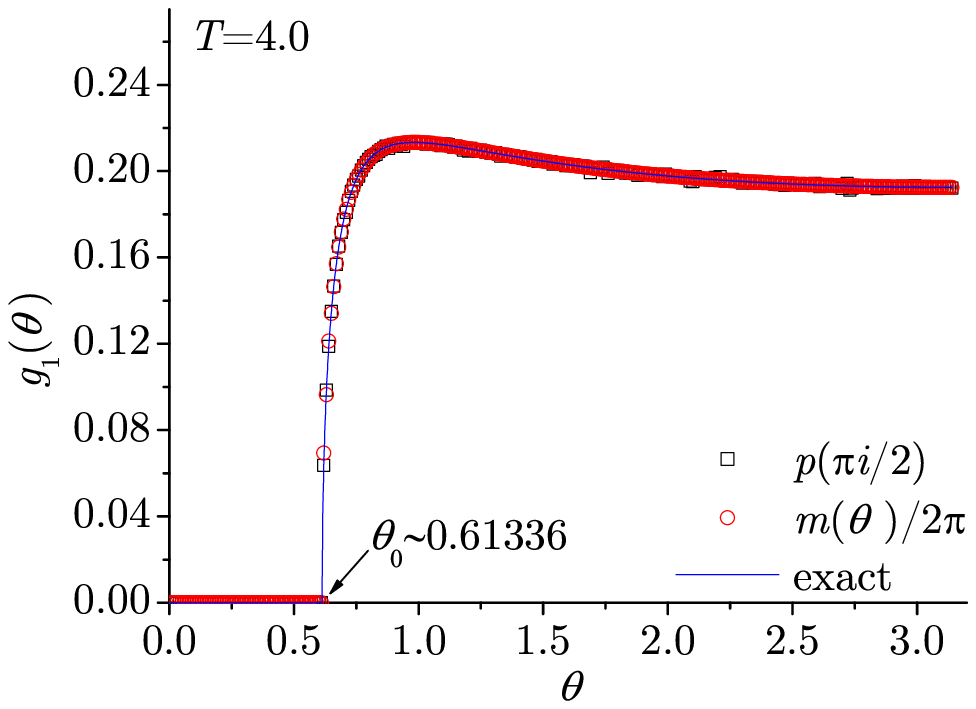}
\end{minipage}
\begin{minipage}{0.5\hsize}
\includegraphics[width=1\linewidth]{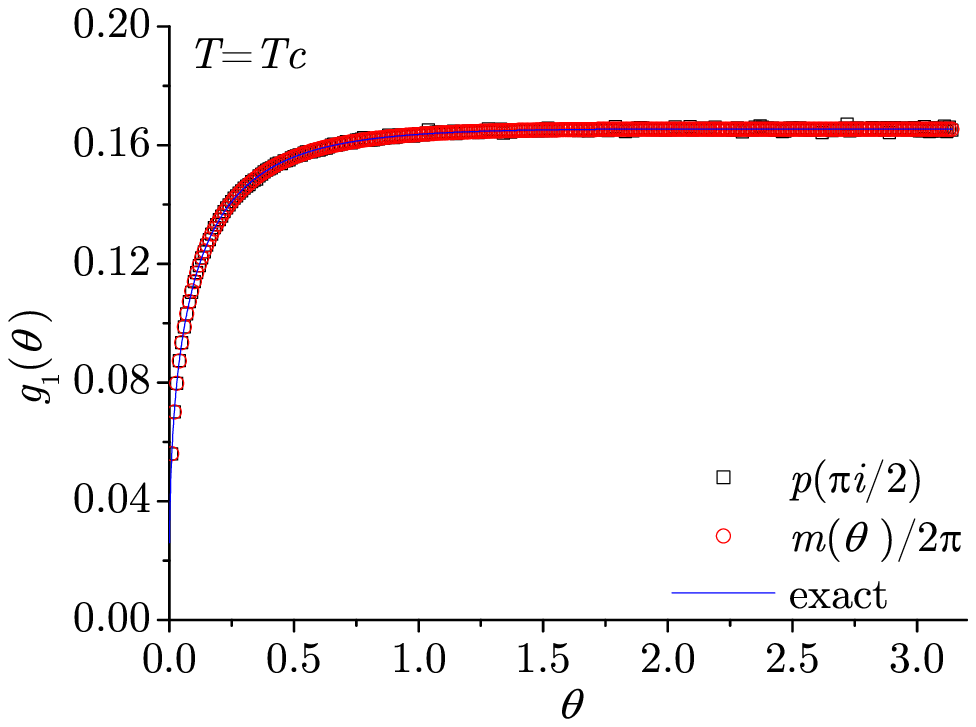}
\end{minipage}
\begin{minipage}{0.5\hsize}
\includegraphics[width=1\linewidth]{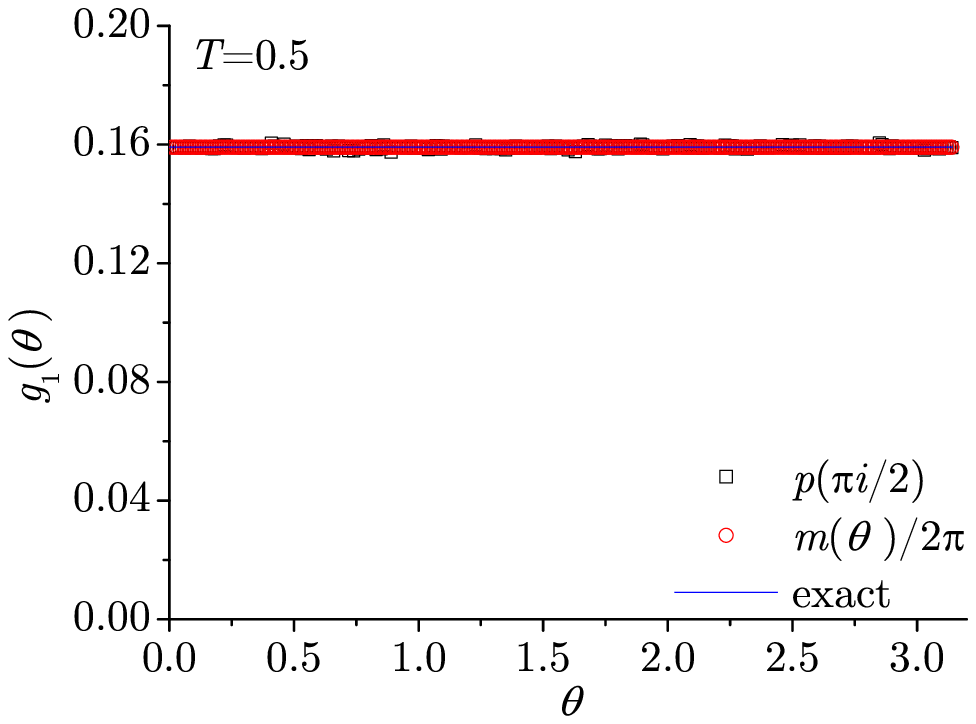}
\end{minipage}
\begin{minipage}{0.5\hsize}
\caption{
The density of zeros on the imaginary axis for the pure ferromagnetic system with $c = 3$. The numerical estimations are consistent with the exact calculation of equation~(\ref{gforf2}). The edge of zeros corresponds to $\theta_0$ in the text.
\label{fig:pureferro}}
\end{minipage}
\end{figure}

In order to check our method itself and the precision of the numerical analyses, we estimated the density of zeros for the pure ferromagnetic system. The zeros of a ferromagnetic system is located only on the imaginary axis, and thus $g_2(H)=0$. In this case the density of zeros on the imaginary axis, $g_1(H)$, can be found by both methods which we used to calculate $g_1$ and $g_2$ in the case $p<1$ (equations~(\ref{eq:PH1}) and (\ref{densityg1})).
 We here compare the exact density of zeros and two algorithms using equations~(\ref{eq:PH1}) and (\ref{densityg1}). Figure \ref{fig:pureferro} is for comparison of numerical calculations and the exact solution. Our algorithms agree perfectly with the exact solution. The details of the analyses are as follows.
\begin{enumerate}
 \item The exact density of zeros (the solid line in figure \ref{fig:pureferro}): The density of zeros on the imaginary field axis $g_1\left( \theta  \right)$ is obtained as the analytic continuation of the real $m(2\beta H)$ from real positive values of $\hat H = 2\beta H$ to purely imaginary $\hat H=i\theta$,
\begin{equation}
2 \pi g_1\left( \theta  \right) = \Re \:  {m\:( {\hat H: = i\theta } )} .
\label{gforf1}
\end{equation}
Using the fixed point $\hat h_f \:( \hat H , \beta )$ of equation~(\ref{recursion1}) with $J_{ij}=1$ and $c = 3$, we rewrite equation~(\ref{gforf1}) as
\begin{equation}
2 \pi g_1\left( \theta  \right) = \Re \left. {\tanh \left[{1 \over 2}\left( {{3 \over 2}\hat h_f (\hat H,\beta ) + \hat H} \right) \right]} \right|_{\hat H := i\theta } .
\label{gforf2}
\end{equation}
From equation~(\ref{recursion1}) the fixed point $\hat h_f \:( \hat H, \beta )$ for the pure ferromagnetic system in real field is found as
\begin{equation}
\hat h_f (\hat H, \beta) = 2\log x',
\end{equation}
where $x'$ is the solutions of the following equation,
\begin{equation}
e^{\hat H} x^3  - e^{\hat H + 2\beta } x^2  + e^{2\beta } x - 1 = 0.
\end{equation}

 \item The relation between density of zeros and cavity fields (squares in figure \ref{fig:pureferro}): The algorithm shown as equation~(\ref{eq:PH1}) connects the density of cavity fields and density of zeros. Since the zeros for the ferromagnetic system lie only on the imaginary field axis, the density $g_1\left( \theta \right)$ is estimated from the distribution of cavity field in a purely-imaginary field. We thus consider the free boundary condition in order to make the cavity field pure imaginary.

For the cavity field in a pure imaginary field $\theta$, the recursion relations (\ref{recursion2}) and (\ref{recursion3}) are rewritten as 
\begin{eqnarray}
\Im \: \hat u_i  &=& 2 \tan ^{ - 1} \left[ {\tanh \beta \tan \left( {\theta/2 + \left(c - 1\right) {\Im \: \hat u_j/2 } } \right)} \right] \:\:\:\: \Re \: \hat u_i = 0, \label{pirr}\\
\Im \: \hat h^{(c)}  &=& \theta + c {\Im \: \hat u_i } \:\:\:\: \Re \: \hat h^{(c)} = 0 ,
\end{eqnarray}
where $\hat u = 2 \beta u$ and $\hat h^{(c)} = 2 \beta h^{(c)}$ and the initial $u$ is set as $u = 0$. We directly estimate the one-dimensional density of the population on the imaginary field axis by using the relation,
\begin{equation}
g_1\left( \theta \right) = { P_H^{(c)} \:( \Im \: \hat h_c = \pi  ) }.
\end{equation}
It can be shown in full generality that this procedure yields a density
$g_1(H)$ which correctly describes the jump of the real part of the
magnetization across the imaginary $H$ axis. 
For the spin-glass model ($p < 1$), this algorithm is used only in the estimation of $g_2$ part (equation~(\ref{density})). Since the boundary condition of the Bethe spin-glass is fixed in order to introduce frustration to the system and the cavity fields are complex, this method is not applicable to the estimation of $g_1$ for the spin-glass model.

 \item The real part of complex magnetization (circles in figure \ref{fig:pureferro}): Complex magnetization in the vicinity of the imaginary field axis is estimated from 
\begin{equation}
m\left( \theta \right) =  \mathop {\lim }\limits_{\hat H_R  \to 0 + } \tanh \left[ \frac{1}{2} \left( {\frac{3}{2}\hat h_f' \left( {\hat H,\beta } \right) + \hat H} \right) \right]_{\hat H = i\theta  + \hat H_R } ,
\end{equation}
where $\hat h_f'$ is the (numerically) fixed point of the complex cavity field. The one-dimensional density on the imaginary axis is also calculated from the real part of the complex magnetization as $g_1 \left( \theta \right) = \Re \: m \left( \theta \right) / 2\pi$. This method is used to calculate $g_1$ for the Bethe spin-glass (equation~(\ref{densityg1})).
\end{enumerate}

\begin{figure}[htbp]
\begin{minipage}{0.5\hsize}
\includegraphics[width=0.98\linewidth]{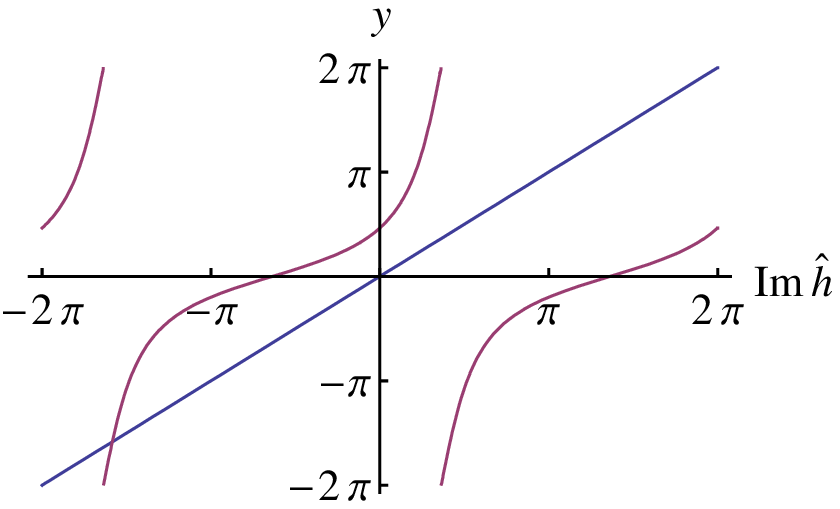}
\end{minipage}
\begin{minipage}{0.5\hsize}
\includegraphics[width=0.98\linewidth]{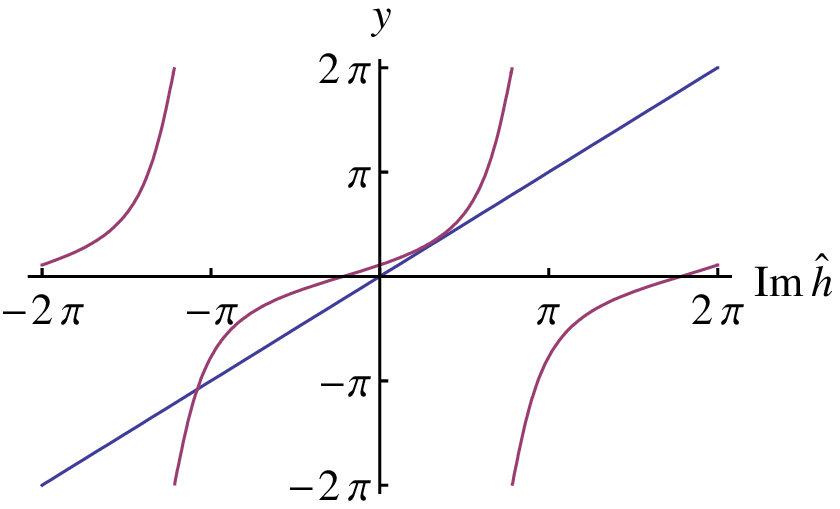}
\end{minipage}
\caption{
The return map of equation~(\ref{recursion3}) for the pure ferromagnetic system in pure imaginary field. Left: If the curve $y = 4 \tan^{ - 1} \left[ {\tanh \beta \tan \left( {\left( {\theta + \Im \hat h} \right)/2} \right)} \right]$ does not intersect $y = \Im \hat h$, the imaginary cavity field does not converge; the density of zeros has a finite value under this condition. Right: If the two equations intersect, the imaginary cavity field recursion has fixed point; thus the density has no value.
\label{fig:returnmap}}
\end{figure}

Note that the imaginary cavity field (equation~(\ref{pirr})) does not converge for the pure ferromagnetic system when the density of zeros has a positive value for a given $\hat H$ (the left panel of figure \ref{fig:returnmap}). However we take the average over a large number of the iterating steps to calculate the density of cavity field, which naturally performs the sampling of bulk properties of the Cayley tree. If the return map of the recursion relation (\ref{recursion1}) has a fixed point, on the other hand, the cavity field population is delta-peaked at that fixed point. One finds that the density of zeros vanishes under this condition (see the right panel of figure \ref{fig:returnmap}).

\begin{figure}[htbp]
\begin{minipage}{0.5\hsize}
\includegraphics[width=1\linewidth]{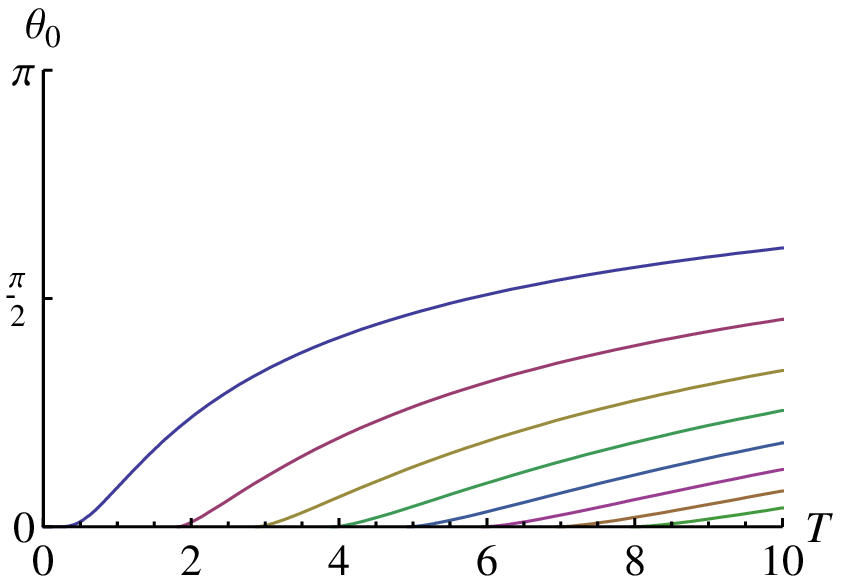}
\end{minipage}
\begin{minipage}{0.5\hsize}
\caption{
The temperature dependence of $\theta_0$ for connectivities $c = 2$, $3$, $\cdots $ $9$ from left to right. At $\theta_0 = 0$, the temperature is given by $T = 1/ \tanh ^{ - 1} \left[ {1/\left( {c - 1} \right)} \right]$.
\label{fig:criticaltheta}}
\end{minipage}
\end{figure}

The critical value of the imaginary field $\theta_0$ is given by the condition that the two curves in figure \ref{fig:returnmap} touch:
\begin{equation}
 2\left( c -1 \right) \tan^{ - 1} \left[ {\tanh \beta \tan \left( \theta_0/2 + \hat h/2 \right)} \right] = \hat h
\end{equation}
and
\begin{equation}
\frac{d}{{d \hat h}} \left[ 2\left( {c - 1} \right)\tan^{ - 1} \left[ {\tanh \beta \tan \left( \theta_0/2 + \hat h/2 \right)} \right] \right] = 1.
\end{equation}
We find that the critical imaginary field  is determined as
\begin{equation}
\begin{array}{r}
\theta _0 \left( \beta  \right) = 2\cos ^{ - 1} \left[ {\sqrt {(c - 1 - \tanh \beta )\sinh \beta \cosh \beta } } \right] \\
- 2\left( {c - 1} \right)\tan ^{ - 1} \left[ \displaystyle{ {\sqrt {\frac{{\left( {1 - \left( {c - 1} \right)\tanh \beta } \right)\tanh \beta }}{{c - 1 - \tanh \beta }}} } } \right].
\end{array}
\end{equation}
The critical temperature is given from this equation at $\theta_0 = 0$ as $\beta  = \tanh ^{ - 1} \left[ {1/\left( {c - 1} \right)} \right]$. This result is of course consistent with the well-known critical temperature. Figure \ref{fig:criticaltheta} shows the temperature dependence of $\theta_0$ for $c = 2$, $3$, $\cdots $ $9$.

\section{
The critical condition based on the spin-glass susceptibility 
}
It is quite possible that 
the divergence of the spin-glass susceptibility $\chi_{SG}$, which
in the SK model is identified with the AT instability \cite{Obuchi:09,Martin:05}, 
characterizes the instability signaled by 
the zeros of the partition function approaching the real axis. Here we show that this is indeed the case for the present Bethe lattice.

The spin-glass susceptibility is defined as
\begin{equation}
\chi_{SG} =\frac{1}{N}\sum_{i,j}\left [
\left(\Part{\Ave{S_{i}}}{h_{j}}{} \right)^2 \right ]_{J}
=\sum_{j}
\left [
\left(\Part{\Ave{S_{0}}}{h_{j}}{} \right)^2 \right ]_{J}
\label{eq:chiSG}.
\end{equation}
To derive the second identity, we assumed uniformity of the Bethe lattice and 
selected the central spin $0$ as $i$.
In a cycle-free graph, an arbitrary pair of sites are connected 
by a single path. Let us assign site indexes from the origin 0 
of the graph to a site of distance $G$ along the path 
as $g=1,2,\ldots,G$. 
For a fixed set of couplings and boundary 
fields, 
the chain rule of the derivative shows that
\begin{eqnarray}
\Part{\Ave{S_{0}}}{h_{G}}{}&=&\Part{\Ave{S_{0}}}{h_{0}}{}
\Part{h_0}{u_0}{}\Part{u_0}{{h}_1}{}
\cdots \Part{h_G}{u_{G}}{}
=\Part{\Ave{S_{0}}}{h_{0}}{}
\Part{h_0}{u_0}{} \prod_{g=1}^G
\Part{u_{g-1}}{h_{g}}{}
\Part{{h}_g}{u_g}{} \cr
&=& \Part{\Ave{S_{0}}}{u_0}{}
\prod_{g=1}^G
\Part{u_{g-1}}{u_g}{},
\label{chainrule}
\end{eqnarray}
as $h_g$ is the cavity field on site $g$ and 
depends linearly on $u_g$ as 
$h_g=H+u_g+r_g$, where $r_g$ represents the sum of 
the cavity biases from the other branches that flow into
site $g$. 
In the limit $G \rightarrow \infty$, the relevant factor in 
equation~(\ref{chainrule}) is only 
$
\prod_{g=1}^G
\left(
\partial u_{g-1}/\partial u_g
\right)
$
. The spin-glass susceptibility is hence evaluated as
\begin{equation}
\chi_{SG}=\sum_{G=1}^{\infty}c(c-1)^{G-1}\left[ \left(\Part{\Ave{S_{0}}}{h_{G}}{}\right)^2
\right]_{J}
\propto 
\sum_{G=1}^{\infty}{
(c-1)^{G}
\left[
\prod_{g=1}^G
\left(
\Part{u_{g-1}}{u_g}{}
\right)^2
\right]_{J}
}
,
\end{equation}
where the factor $c(c-1)^{G}$ denotes the number of sites of distance $G$ 
from the central site $0$.
The divergence condition of $\chi_{SG}$ is given by 
\begin{equation}
\log(c-1)+\lim_{G\to \infty} \frac{1}{G}\log\left(
\left[
\prod_{g=1}^G
\left(
\Part{u_{g-1}}{u_g}{}
\right)^2
\right]_{J}
\right)=0.\label{eq:AT}
\end{equation}
This yields the condition of the spin-glass transition of the Bethe lattice and RRG.

In order to estimate the divergence points of the spin-glass susceptibility, 
we numerically implement the calculation of 
the factor 
$
\left[\prod_{g=1}^G
\left(
\partial u_{g-1}/\partial u_g
\right)^2
\right]_{J}
$.
This factor can also be written as 
$[(\partial u_{0}/\partial u_G)^2]_{J}$ 
and this latter form is more tractable 
at finite temperatures.

The numerical evaluation of the factor $\partial u_{0}/\partial u_G$ is 
straightforward,
\begin{eqnarray}
\Part{u_0}{u_G}{} &\approx&  
\frac{{u_0 \left( {u_G  + \Delta u_G } \right) - u_0
 \left( {u_G } \right)}} {\Delta u_G }.
\label{eq:ATnum}
\end{eqnarray}
The procedure to evaluate this equation is as follows~\cite{Krzakala:05,Rivoire:03,Zdeborov'a:09}. 
We arrange two replicas of 
an identical population $\{u_{i}\}_{i=1}^{N_{\rm pop}}$ 
expressing
the convergent (real) cavity-bias distribution $\pi_{H,\beta}\left( u \right)$, which 
is related to the convergent cavity-field distribution $P_{H,\beta}(h)$ given in 
equation~(\ref{eq:convergentCFD}) as 
\begin{eqnarray}
\pi_{H,\beta}(u)=\int 
dh P_{H,\beta}(h)
\left[
\delta\left(
u-\frac{1}{\beta}\tanh^{-1}
\left\{
\tanh \beta J 
\tanh \beta h 
\right\}
\right)
\right]_{J}.\label{eq:CFDtoCBD}
\end{eqnarray}
In addition, we introduce 
a uniform perturbation ($\Delta u=10^{-4}$) 
into only one of the two replicas 
and then observe the square average of the variation, 
$(1/N_{\rm pop})\sum_{i=1}^{N_{\rm pop}}(u_{i}(u_G  + \Delta u_G) - u_{i}(u_G))^2$, 
after a certain number of the cavity updates.

\begin{figure}[htbp]
\begin{minipage}{0.5\hsize}
\includegraphics[width=1\linewidth]{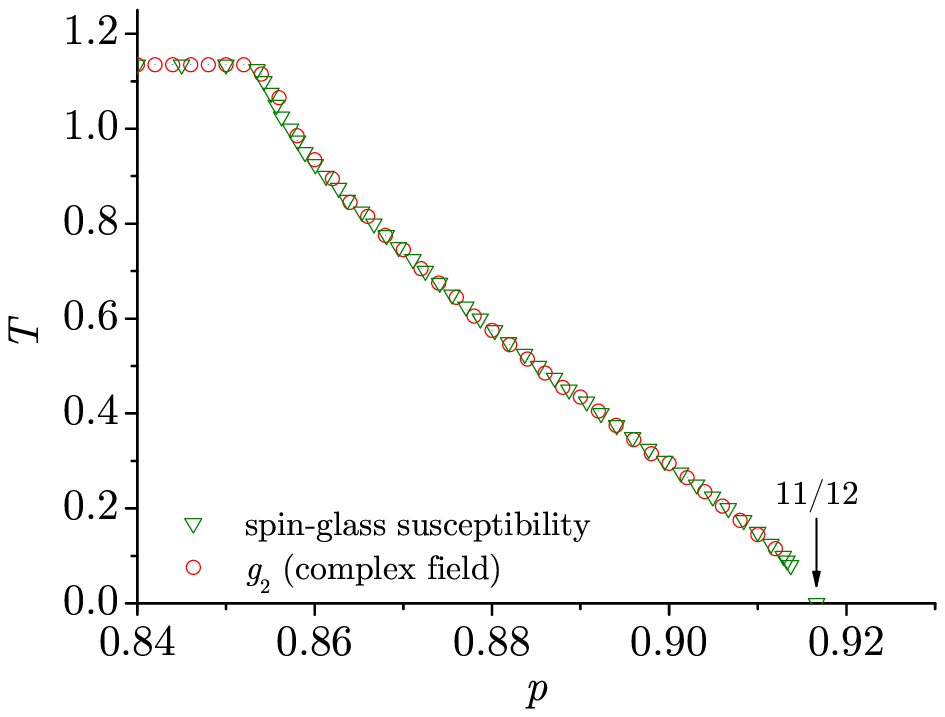}
\end{minipage}
\begin{minipage}{0.5\hsize}
\caption{
The divergence points of the spin-glass susceptibility calculated by the instability of the real cavity field distribution. The critical probability at zero temperature is suggested as $p_c = 0.91665(5)$. The circles denote the phase boundary estimated by the two-dimensional distribution of zeros (figure \ref{fig:pdpt}). They agree well with each other.
\label{fig:sgsus}}
\end{minipage}
\end{figure}

In particular, 
we update two populations by $5000N_{\rm pop}$ iterations with the same set of $J_{ij}$. A critical line of the divergence of the spin-glass susceptibility is determined by whether the square average is numerically zero or much larger than the perturbation. The result is shown in figure \ref{fig:sgsus} where the spin-glass susceptibility diverges below this line. At zero temperature, the critical probability is calculated as $p_c = 11/12$ in references~\cite{Castellani:05,Kwon:88}, which is reproduced by our numerical calculation at zero temperature as $p_c = 0.91665(5)$.

This result agrees well with the phase boundary drawn in figure \ref{fig:pdpt}. Thus it is suggested that our phase boundary estimated by the two-dimensional zeros corresponds to the spin-glass transitions.

\section*{References}

\providecommand{\newblock}{}

\end{document}